\documentclass[a4paper,11pt]{article}
\pdfoutput=1
\usepackage{jcappub}

\usepackage{amsmath,amssymb,graphicx}
\usepackage{soul}
\usepackage[latin1]{inputenc}
\usepackage{dsfont}

\usepackage{subfigure}

\usepackage{array}
\usepackage{mathtools}

\usepackage[thinlines]{easytable}

\newcommand{\be}{\begin{eqnarray}}
\newcommand{\en}{\end{eqnarray}}
\newcommand{\nn}{\nonumber\\}
\newcommand{\Hubble}{ H }

\newcommand{\mpl}{m_{\rm{pl}}}

\newcommand{\braket}[1]{\langle #1 \rangle}

\usepackage{afterpage}

\begin{document}

\begin{titlepage}

\vspace*{-15mm}
\vspace*{0.7cm}

\begin{center}

{\Large {\bf Hill crossing during preheating after hilltop inflation}}\\[8mm]

Stefan Antusch$^{\star\dagger}$\footnote{Email: \texttt{stefan.antusch@unibas.ch}},  
David Nolde$^{\star}$\footnote{Email: \texttt{david.nolde@unibas.ch}} and 
Stefano Orani$^{\star}$\footnote{Email: \texttt{stefano.orani@unibas.ch}}

\end{center}

\vspace*{0.20cm}

\centerline{$^{\star}$ \it
Department of Physics, University of Basel,}
\centerline{\it
Klingelbergstr.\ 82, CH-4056 Basel, Switzerland}

\vspace*{0.4cm}

\centerline{$^{\dagger}$ \it
Max-Planck-Institut f\"ur Physik (Werner-Heisenberg-Institut),}
\centerline{\it
F\"ohringer Ring 6, D-80805 M\"unchen, Germany}

\vspace*{1.2cm}

\begin{abstract}
\noindent
In ``hilltop inflation'', inflation takes place when the inflaton field slowly rolls from close to a maximum of its potential (i.e.\ the ``hilltop'') towards its minimum. When the inflaton potential is associated with a phase transition, possible topological defects produced during this phase transition, such as domain walls, are efficiently diluted during inflation. It is typically assumed that they also do not reform after inflation, i.e.\ that the inflaton field stays on its side of the ``hill'', finally performing damped oscillations around the minimum of the potential. In this paper we study the linear and the non-linear phases of preheating after hilltop inflation. We find that the fluctuations of the inflaton field during the tachyonic oscillation phase grow strong enough to allow the inflaton field to form regions in position space where it crosses ``over the top of the hill'' towards the ``wrong vacuum''. We investigate the formation and behaviour of these overshooting regions using lattice simulations: Rather than durable domain walls, these regions form oscillon-like structures (i.e.\ localized bubbles that oscillate between the two vacua) which should be included in a careful study of preheating in hilltop inflation.
\end{abstract}
\end{titlepage}

\tableofcontents

\section{Introduction}

Inflation has proven to be a very successful paradigm in early universe cosmology. It successfully solves the horizon and flatness problems of Big Bang cosmology and at the same time provides an explanation for the adiabatic, nearly scale invariant and Gaussian primordial curvature perturbations implied by observations of the cosmic microwave background and large scale structure \cite{Ade:2015lrj,Ade:2015tva,Beutler:2011hx,Anderson:2013zyy,Ross:2014qpa}.

At the end of inflation, the universe is dominated by the inflaton field's potential energy which is converted into radiation in a process called reheating. Reheating typically happens in at least two stages: an initial ``preheating'' phase of non-perturbative particle production, and a later stage of perturbative inflaton decay and thermalization. Together, these processes describe the transition of the inflationary, vacuum dominated phase to the radiation dominated universe of standard $\Lambda$CDM cosmology. A detailed understanding of the reheating phase is required not only for precise predictions of the primordial spectrum (as the expansion history after inflation affects the matching of inflationary and CMB perturbations), but also for calculating the non-thermal production of e.g.\ a baryon asymmetry, dark matter or other relics from inflaton decays.

Despite its success in cosmology, it is not yet clear how exactly inflation is realized in terms of particle physics. One of the many interesting possibilities is hilltop inflation, also called ``new inflation'' (see e.g.\ \cite{Linde:1981mu,Izawa:1996dv,Izawa:1997df,Senoguz:2004ky,Boubekeur:2005zm,Kohri:2007gq}): the inflaton $\phi$ starts at a local maximum of the scalar potential at $\phi=0$ and slowly rolls towards a minimum of the scalar potential at $\phi \neq 0$, spontaneously breaking some particle physics symmetry like a family symmetry \cite{Antusch:2008gw}. The initial conditions for hilltop inflation ($\phi$ starts near a local maximum) can be easily generated by a coupling of the inflaton to some matter field $\chi$, which allows for a period of preinflation \cite{Izawa:1997df,Senoguz:2004ky,Yamaguchi:2004tn,Antusch:2014qqa} and also provides an efficient decay channel for reheating \cite{Antusch:2014qqa}. Other proposals for generating the initial conditions include \cite{Vilenkin:1983xq,Vilenkin:1994pv,Guth:1985ya,Ellis:2014rja}.

Preheating in hilltop inflation has been studied numerically in \cite{Desroche:2005yt} and analytically in \cite{Brax:2010ai}. Both papers find that perturbations are amplified by both tachyonic preheating and parametric resonance, and that perturbations grow very large at scales corresponding to the inflaton mass. Detailed analytical calculations for the linearised perturbations are presented in \cite{Brax:2010ai}; however, when the perturbations eventually grow too large, the linear approximation breaks down and any further calculation must include the non-linear interactions. \cite{Desroche:2005yt} studies preheating numerically using lattice simulations, using a slightly different (logarithmic) hilltop potential. They study the evolution of particle number densities $n_k$ and find that the growth of perturbations is terminated by the non-linear interactions when $\braket{\delta \phi^2} \sim \braket{\phi}^2$.

In this paper, we study both the linear and the early non-linear stages of preheating in hilltop inflation. Beyond confirming the earlier results of \cite{Desroche:2005yt} and \cite{Brax:2010ai}, we study the evolution of the inhomogeneous inflaton field in position space. We are particularly interested in the possibility that the large inflaton fluctuations might push the inflaton field over the local maximum of the scalar potential towards the ``wrong'' vacuum, which could in principle result in the formation of domain walls. We find that such ``hill crossing'' indeed occurs for a wide range of parameters. However, it turns out that these regions of ``wrong'' vacuum do not form domain wall networks, but that they instead generate localized oscillating bubbles, which might be interpreted as oscillons \cite{Copeland:1995fq,Adib:2002ff,Gleiser:2003uu,Graham:2006xs,Farhi:2007wj,Amin:2011hj,Zhou:2013tsa,Gleiser:2014ipa,Lozanov:2014zfa}.

The paper is structured as follows. In section~\ref{sec:hilltopModel}, we introduce the hilltop inflation model studied in this paper. The linear phase of preheating is discussed in section~\ref{sec:linear} based on \cite{Brax:2010ai}, with special emphasis on the exponential growth of perturbations which suggests that perturbations might grow large enough to push the inflaton field over the hilltop in some regions of space. However, the linear approximation breaks down for such large perturbations, so to check whether hill crossing can actually happen, we study the non-linear preheating dynamics using lattice simulations in section~\ref{sec:lattice}. We then summarize our results in section~\ref{sec:summary}.

\section{Hilltop inflation}
\label{sec:hilltopModel}

\begin{figure}[tbp]
  \centering
\includegraphics[width=0.9\textwidth]{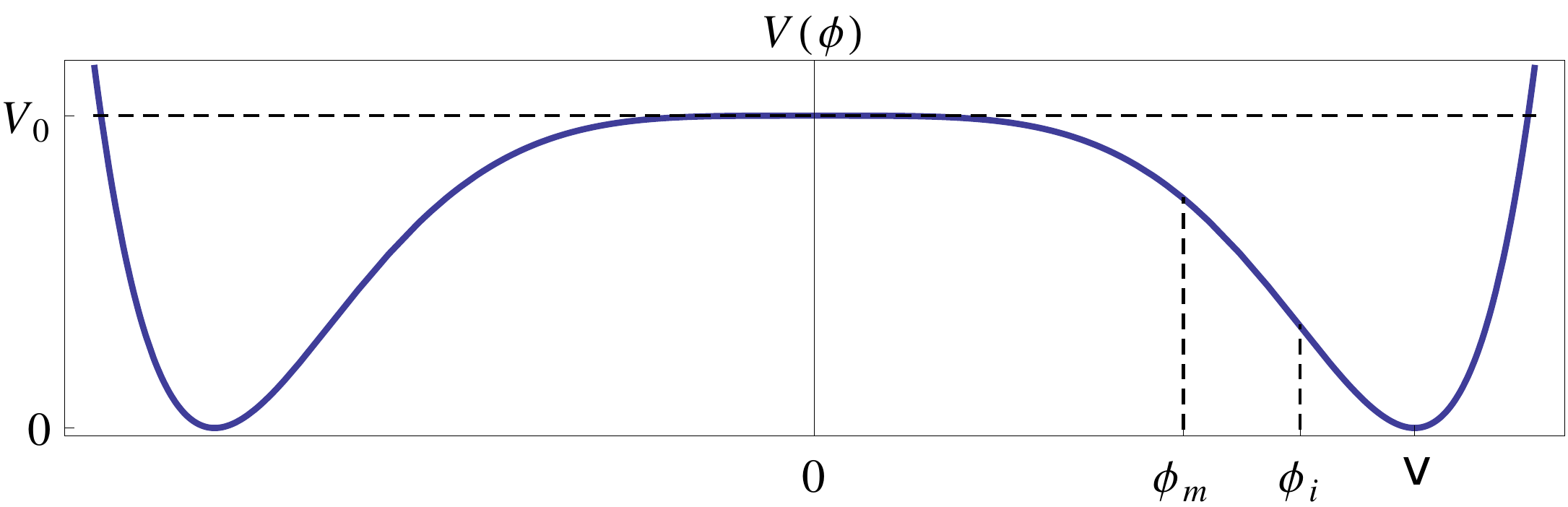}
  \caption{Inflaton potential with the points of maximum tachyonic mass ($\phi_{\rm m}$), the end of tachyonic preheating ($\phi_{\rm i}$) and the potential minimum ($v$). Inflation happens at the hilltop for $\lvert \phi \rvert \lesssim v^2/(\sqrt{24}\mpl)$ as the inflaton field slowly rolls down towards the minimum at $\phi = v$.}
  \label{fig:potential}
\end{figure}

In this paper, we consider the hilltop inflation potential
\begin{align}
 V(\phi) \, = \, V_0 \left( 1 - \frac{\phi^4}{v^4} \right)^2 \label{eq:potential}
\end{align}
with $v \ll \mpl$. This potential is plotted in fig.~\ref{fig:potential}.

Hilltop inflation happens near $\phi=0$, with $\phi$ slowly rolling down the hilltop, while the universe expands by at least $N_* \sim 60$ $e$-folds to solve the horizon and flatness problems. We choose $\phi>0$, which can always be arranged by a field redefinition $\phi \rightarrow -\phi$. Slow-roll inflation ends shortly after $\eta = \mpl^2 V''/V = -1$, which happens at the inflaton field value $\lvert \phi_\eta \rvert \simeq v^2/\sqrt{24}$, and the inflaton quickly rolls towards the minimum at $\phi = v$.\footnote{In fact, inflation continues for $\sim 2.5$ $e$-folds after $\phi_\eta$, but the expansion in potential slow-roll parameters breaks down around $\eta \sim -1$.}

\subsubsection*{Primordial perturbations}
The scalar perturbation amplitude matches the observed value $A_s = 2.2 \times 10^{-9}$ for
\begin{align}
 V_0 \simeq 2\times 10^{-14} \, v^4,
\end{align}
and the predictions for the spectral index $n_s$ and tensor-to-scalar ratio $r$ are
\begin{align}
n_s = 1-\frac{3}{N_*} \simeq 0.95, \quad\quad r \simeq 6 \times 10^{-7} \left( \frac{v}{\mpl} \right)^4,
\end{align}
in some tension with the most recent Planck bounds $n_s = 0.965 \pm 0.005$ (at 68\% CL) \cite{Ade:2015lrj}. The predicted value of $n_s$ can be larger if there is an additional tachyonic mass term for $\phi$, which occurs naturally in supergravity implementations of hilltop inflation due to corrections from the K\"{a}hler potential, or if the $\phi^4/v^4$ in the inflaton potential is replaced by some higher power $\phi^p/v^p$ with $p>4$.

\subsubsection*{Initial conditions from preinflation}
The initial conditions for hilltop inflation ($\phi \simeq 0$) can be naturally generated e.g.\ by a phase of preinflation \cite{Izawa:1997df,Senoguz:2004ky,Yamaguchi:2004tn,Antusch:2014qqa}. During preinflation, another scalar field $\chi$ slowly rolls down its potential, while $\phi$ is driven to zero by a large mass term generated from couplings like $V_{\rm int} = \lambda \phi^2 \chi^4$ \cite{Antusch:2014qqa}. Eventually, when $\chi$ reaches very small values, the potential for $\phi$ takes the form of eq.~\eqref{eq:potential}, and $\phi$ starts to slow-roll away from zero either towards $\phi=v$ or $\phi=-v$, starting with an initial displacement from $\phi=0$ due to quantum fluctuations. This second slow-roll phase, which can be described as single-field hilltop inflation in $\phi$ with the potential \eqref{eq:potential}, lasts more than $N_* \sim 60$ $e$-folds. As a consequence, any topological defects generated during the initial symmetry breaking are inflated away, and at the end of inflation the entire observable universe has either $\phi > 0$ or $\phi < 0$.

\subsubsection*{Preheating}
Preheating in hilltop inflation has been shown to happen in two steps \cite{Brax:2010ai}: tachyonic preheating and tachyonic oscillations.

\begin{enumerate}
 \item While the inflaton rolls from $\phi_\eta$ to $\phi_{\rm i} = \left(3/7\right)^{1/4} v \simeq 0.81 v$, the inflaton mass is tachyonic: $V''(\phi) < 0$. During this phase, long-wavelength perturbations with $k \lesssim H_0$ grow exponentially in a process called ``tachyonic preheating''.
\item For $v \gtrsim 10^{-5}\mpl$, $\phi$ does not quickly relax at the minimum. Instead, it oscillates between $\phi \simeq 0$ and $\phi \simeq 2^{1/4}v$. These ``tachyonic oscillations'' cause a strong growth in small-wavelength perturbations around a scale $k \sim k_{\rm peak} \simeq 7 (\mpl/v)^{3/4} H_0$, which grow until the perturbations become non-linear (for $v/\mpl \lesssim 10^{-2}$) or until Hubble damping reduces the oscillation amplitude so much that the inflaton oscillations remain in the approximately quadratic part around the minimum.
\end{enumerate}

\section{Linear preheating}
\label{sec:linear}
In this section, we discuss the phase of linear preheating after inflation. During this phase, the inflaton field is still almost homogeneous with small fluctuations $\delta \phi$ which can be treated as small perturbations (keeping only terms to leading order in $\delta \phi$). The initial phase of preheating can be studied by solving the time evolution of these fluctuations in Fourier space.

We start with a brief review of the theory of linear preheating and its application to the hilltop inflation potentials in eq.~\eqref{eq:potential}. For these potentials, preheating starts with tachyonic preheating, followed by a period of tachyonic oscillations. We summarize the analytic results from \cite{Brax:2010ai} for these two phases and show numerical results for the spectra produced by linear preheating. We then discuss how these power spectra suggest that for $10^{-5} \lesssim v/\mpl \lesssim 10^{-2}$, the perturbations could push the inflaton field above the hilltop towards $\phi(x) < 0$, a process which could potentially produce domain walls.

\subsection{Introduction to linear preheating}
\subsubsection{Equations of motion}
The equation of motion of the inflaton field $\phi(t,\vec{x})$ in a FRW background (neglecting backreaction from the metric perturbations\footnote{To account for scalar metric perturbations, one can use the Mukhanov-Sasaki variable instead of the inflaton field value \cite{Sasaki:1986hm,Mukhanov:1988jd}. For hilltop inflation, the effect on the tachyonic preheating spectrum is known to be negligible \cite{Brax:2010ai}. We have also checked for a few parameter points that including metric perturbations in this way did not make a noticeable difference for the subsequent oscillation phase, so analogously to \cite{Brax:2010ai}, we perform our calculations using the scalar field eq.~\eqref{eq:eomPhi}, neglecting metric perturbations.}) is given by
\begin{align}
 \ddot{\phi} \, + \, 3\Hubble \dot{\phi} \, - \, \vec{\nabla}^2 \phi \, + \, V'(\phi) \, = \, 0. \label{eq:eomPhi}
\end{align}
At the end of inflation, the inflaton field is almost homogeneous, so we can expand it in terms of a homogeneous mean field $\overline{\phi}(t)$ and a perturbation field $\delta \phi(t, \vec{x})$:
\begin{align}
 \phi(t,\vec{x}) \, = \, \overline{\phi}(t) + \delta \phi(t, \vec{x}).
\end{align}
As long as the perturbations remain small, we can linearise eq.~\eqref{eq:eomPhi} by dropping all terms of $\mathcal{O}(\delta \phi^2)$:
\begin{subequations}
\begin{align}
  \ddot{\overline{\phi}} \, + \, 3\Hubble \hspace{0.3pt} \dot{\overline{\phi}} \, + V'(\overline{\phi}) \, = \, 0,
\end{align}
\begin{align}
 \delta \ddot{\phi} \, + \, 3\Hubble \hspace{0.3pt} \delta \dot{\phi} \, + \left( V''(\overline{\phi}) - \, \vec{\nabla}^2 \right) \delta \phi \, = \, 0. \label{eq:eomDeltaPhi}
\end{align}
\end{subequations}
The partial differential equation~\eqref{eq:eomDeltaPhi} for $\delta \phi(t,\vec{x})$ can be transformed into ordinary differential equations by transforming to Fourier space:
\begin{align}
 \ddot{\phi}_k \, + \, 3\Hubble \dot{\phi}_k \, + \left( V''(\overline{\phi}) + \, \frac{k^2}{a^2} \right) \phi_k \, = \, 0, \label{eq:eomDeltaPhiK}
\end{align}
where $\phi_k$ is the Fourier mode with comoving momentum $k$.

\subsubsection{Initial conditions from Bunch-Davies vacuum}
The initial conditions are given by matching to the Bunch-Davies vacuum at early times:
\begin{align}
 \lim\limits_{\mathclap{t \rightarrow - \infty}} \, \phi_k(t) \, \simeq \, \frac{\Hubble}{\sqrt{2k^3}} \left(  i + \frac{k}{a H}  \right) e^{i k / (a \Hubble)}. \label{eq:initialBD}
\end{align}
For a numerical implementation, where matching at $t \rightarrow - \infty$ is not practical, ``early times'' means that the Hubble constant at the time of matching must still be approximately constant, and that the inflaton mass $V''(\overline{\phi})$ should be negligible compared to $k^2/a^2$.

\subsubsection{Inflaton fluctuation amplitude and power spectrum}
Given eqs.~\eqref{eq:eomDeltaPhiK} and \eqref{eq:initialBD}, one can calculate the evolution of the scalar field perturbations for each mode $k$. The result can be translated into position space perturbations by an inverse Fourier transformation. For the inflaton field's variance, this gives
\begin{align}
 \braket{\delta \phi^2(t,\vec{x})} \, = \, \int\limits \left( d \ln k \right) \mathcal{P}_\phi(t, k),
\end{align}
where the limits of the integral are the IR and UV cutoffs,\footnote{The infrared cutoff can be naturally introduced by considering a box of finite size, e.g.\ a single Hubble patch or the observable universe. The ultraviolet cutoff must be put in by hand to regularize the usual UV divergence arising from small-scale quantum fluctuations. This cutoff should be chosen large enough to encompass the largest $\phi_k$ amplified by preheating dynamics, but small enough for the non-amplified quantum fluctuations to be subdominant.} and for $d=3$ spatial dimensions the spectrum $\mathcal{P}_\phi$ can be calculated as
\begin{align}
 \mathcal{P}_\phi (t, k) \, = \, \frac{ k^3 }{ 2\pi^2 } \left| \phi_k(t) \right|^2.
\end{align}

The results from linear preheating are reliable only until higher orders in $\delta \phi$ become important. At that point, the fully non-linear eq.~\eqref{eq:eomPhi} must be solved. This is usually done by solving the equation numerically on a discrete spacetime lattice, as we will discuss in section~\ref{sec:lattice}. For the hilltop potential \eqref{eq:potential}, we can estimate that the linear approximation is valid approximately until $\braket{\delta \phi^2} \gtrsim \overline{\phi}\vphantom{\phi}^2$, or equivalently $\max\limits_k \mathcal{P}_\phi \gtrsim \overline{\phi}\vphantom{\phi}^2$.

\subsection{Tachyonic preheating}
When $V''(\phi) < 0$, all modes with $k/a < \sqrt{-V''(\phi)}$ experience exponential growth. This period of tachyonic growth lasts the longest for modes with $k \lesssim H_0$, which are already tachyonic at the end of inflation. Higher $k$ modes become tachyonic later, and therefore achieve a smaller final amplitude because they grow for a shorter time. Modes above $k_{\rm m}/a = \max\limits_{\phi} \sqrt{- V''(\phi)} \simeq 4 (\mpl/v)H_0 $ are not amplified at all.

After tachyonic preheating\footnote{The value of the power spectrum in eq.~\eqref{eq:tachyonicPk} is evaluated at $\phi = v$ at which time the tachyonic peak takes on its maximum.}, the power spectrum around the Hubble scale is
\begin{align}
\label{eq:tachyonicPk}
 \mathcal{P}_\phi(aH_0) \, \sim \, 10^{-13} \,\mpl^2,
\end{align}
independent of $v$. For $aH_0 < k \lesssim k_{\rm m}$, the power spectrum falls off as $\mathcal{P}_\phi \propto (k/a)^{-2}$, and for $k \gtrsim k_{\rm m}$, the spectrum is just the vacuum spectrum $\mathcal{P}_\phi \simeq k^2/(4\pi^2 a^2)$.

The variance of the inflaton field is dominated by the large contribution from the Hubble scale:
\begin{align}
\label{eq:tachyonicVariance}
 \braket{ \delta \phi^2(x) } \, = \, \int \left( d \log k \right) \mathcal{P}_\phi(k) \, \sim \, \mathcal{P}_\phi(aH_0) \, \sim 10^{-13}\,\mpl^2.
\end{align}
For very small $v/\mpl \lesssim 10^{-6}$, this implies $\braket{ \delta \phi^2(x) } \gtrsim v^2$, so the perturbations become non-linear during tachyonic preheating -- the inflaton field becomes very inhomogeneous and the description in terms of a background field plus perturbations breaks down. For larger $v/\mpl \gtrsim 10^{-5}$, eq.~\eqref{eq:tachyonicVariance} implies $\braket{ \delta \phi^2 } \ll v^2$ and the linear approximation should remain viable.

\subsection{Tachyonic oscillations}
\begin{figure}[ptb]
  \centering$
\begin{array}{cc}
\includegraphics[width=0.48\textwidth]{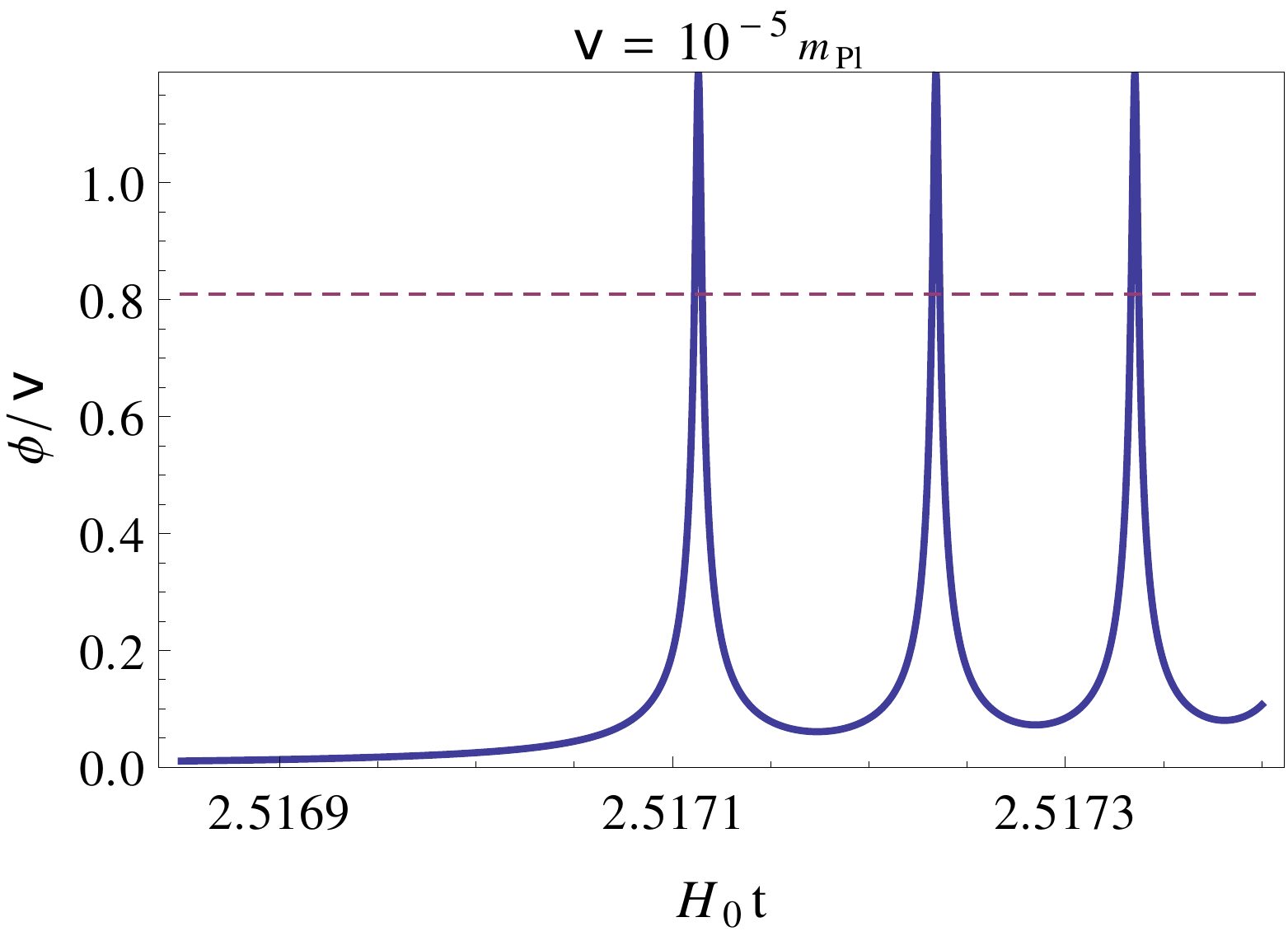} &
\includegraphics[width=0.48\textwidth]{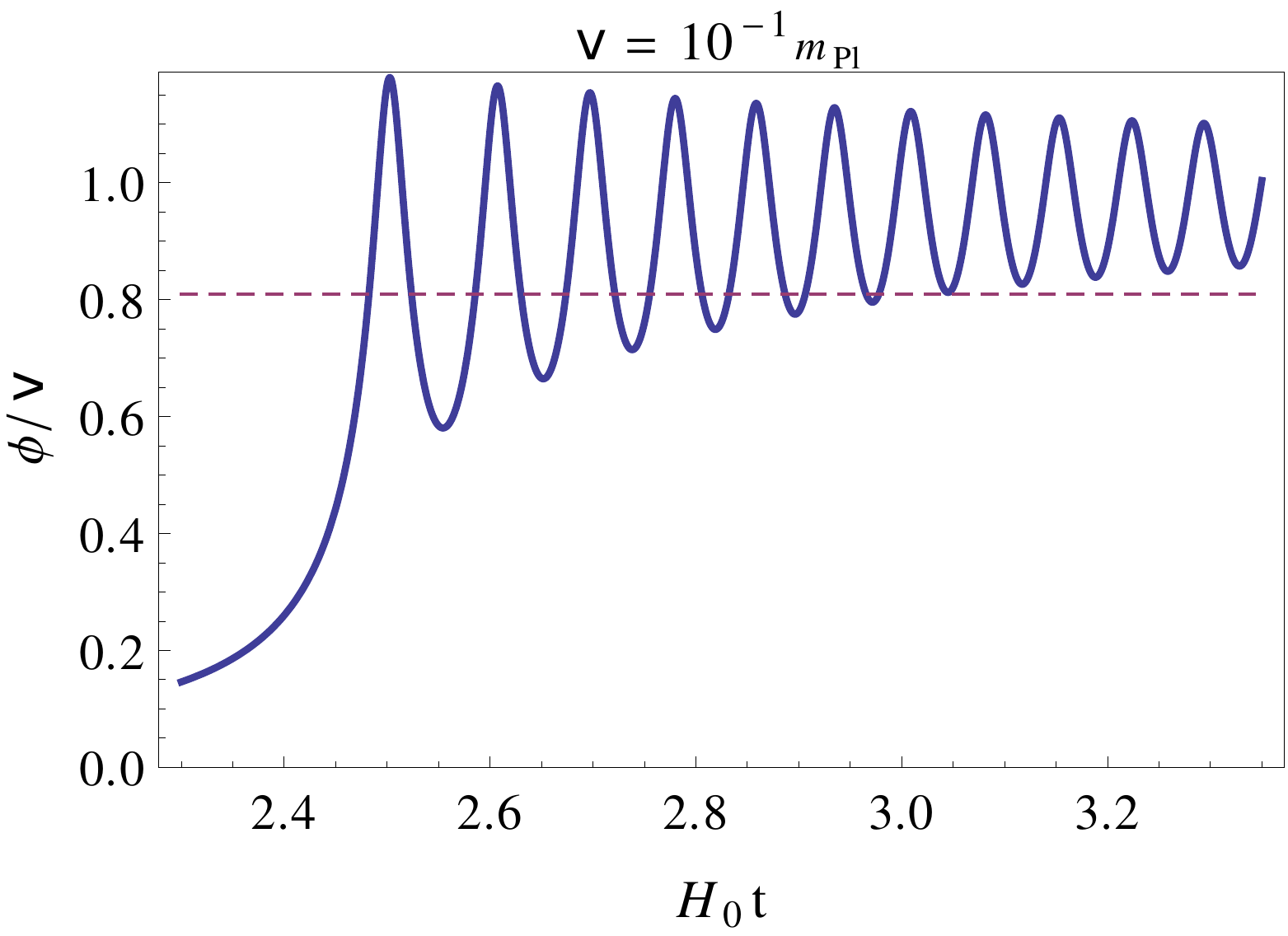}
\end{array}$
  \caption{Oscillations of the inflaton background field around the minimum as a function of $H_0 t$, plotted for $v/\mpl=10^{-5}$ and $v/\mpl=10^{-1}$. The dashed horizontal line denotes the inflaton value $\phi_{\rm i} \simeq 0.81 v$ below which $V''(\phi)<0$. We see that for small $v/\mpl \ll 10^{-1}$, the oscillations continue to bring $\phi$ deeply into the tachyonic region for several oscillations, which explains why the perturbations grow non-linear within a few oscillations. For large $v/\mpl=10^{-1}$, however, Hubble damping quickly dampens the oscillation amplitude so that after a few oscillations, the inflaton field never re-enters the tachyonic region. Preheating is much less efficient in this case: the perturbations remain small and are eventually redshifted away as the inflaton field oscillates around the quadratic part of its potential.}
  \label{fig:inflatonOscillations}
\end{figure}

During the subsequent phase of tachyonic oscillations, the background field oscillates around the minimum. For $v/\mpl \lesssim 10^{-1}$, these oscillations take the inflaton back into the tachyonic region with $\phi < \phi_{\rm i}$. In this case, one can estimate the inflaton field value after $j$ complete oscillations as
\begin{align}
 \frac{\phi_j}{v} \, \simeq \, \left( \sqrt{3} j \frac{v}{\mpl} \right)^{1/4}.
\end{align}
These oscillations trigger strong growth of perturbations at a characteristic peak scale
\begin{align}
 k_{\rm peak} \, \simeq \, 4\sqrt{2} j^{1/4} \left( \frac{\mpl}{v} \right)^{3/4} aH_0.
\end{align}
For small $v/\mpl \ll 10^{-1}$, the perturbations grow by many orders of magnitude and become non-linear within a few oscillations around the minimum. For large $v/\mpl \gtrsim 10^{-1}$, however, the oscillations are quickly damped due to Hubble damping (see fig.~\ref{fig:inflatonOscillations}). After a few oscillations, the amplitude decreases so much that the inflaton field never re-enters the tachyonic region, which makes the preheating much less efficient. In this case, the perturbations remain so small that we expect the linear approximation to remain valid throughout the entire preheating phase.\footnote{There is some non-adiabatic growth of the inflaton perturbations during the subsequent non-tachyonic oscillations around the minimum, but red-shifting and Hubble damping make the perturbations decay before they can grow non-linear.}

Note that the perturbations with $k \ll k_{\rm peak}$, which are amplified most during tachyonic preheating, do not continue to grow during the oscillation phase. Even though they are amplified every time that the inflaton field rolls down the tachyonic part of the potential, this amplification is exactly cancelled by an exponential damping sustained every time the field rolls back up. The infrared part of the spectrum therefore oscillates: it is very large near the minimum at $\phi \sim v$, and it is negligible at $\phi_j$ (i.e.\ near the hilltop).

\subsection{Numerical results for the power spectra}
\label{sec:linearPowerSpectra}

\begin{figure}[ptb]
  \centering$
\begin{array}{cc}
\includegraphics[width=0.48\textwidth]{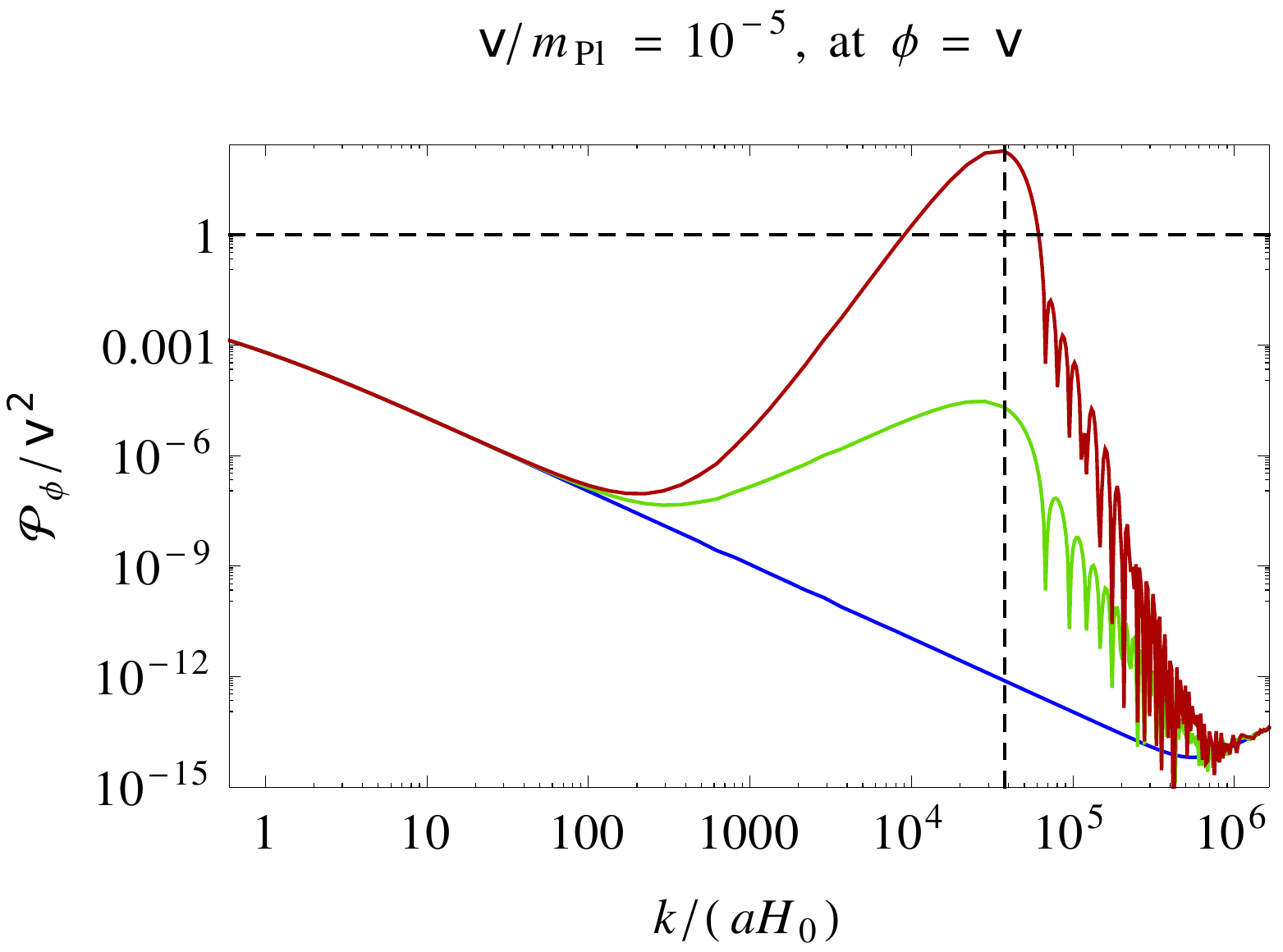} &
\includegraphics[width=0.48\textwidth]{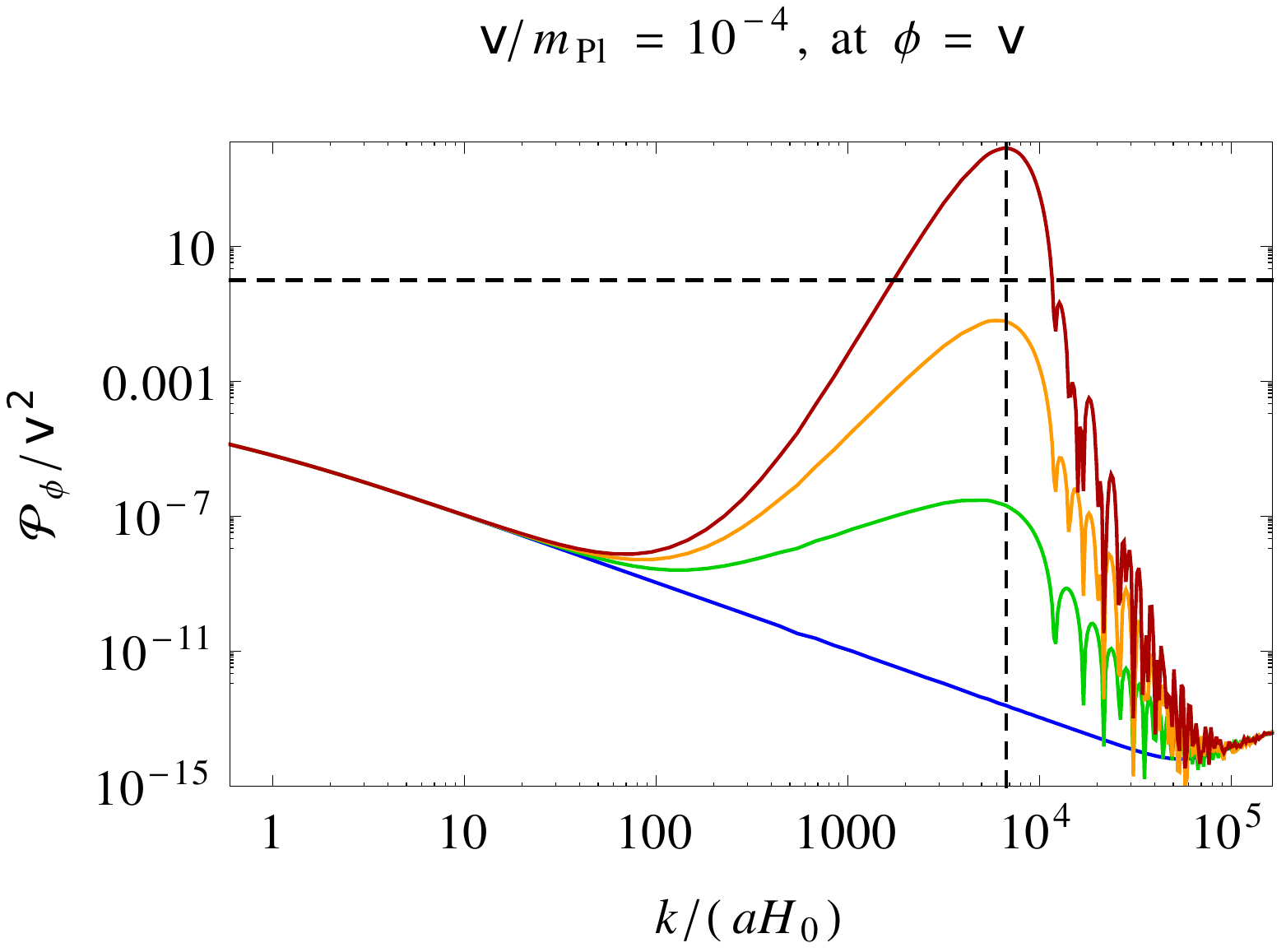} \\
\includegraphics[width=0.48\textwidth]{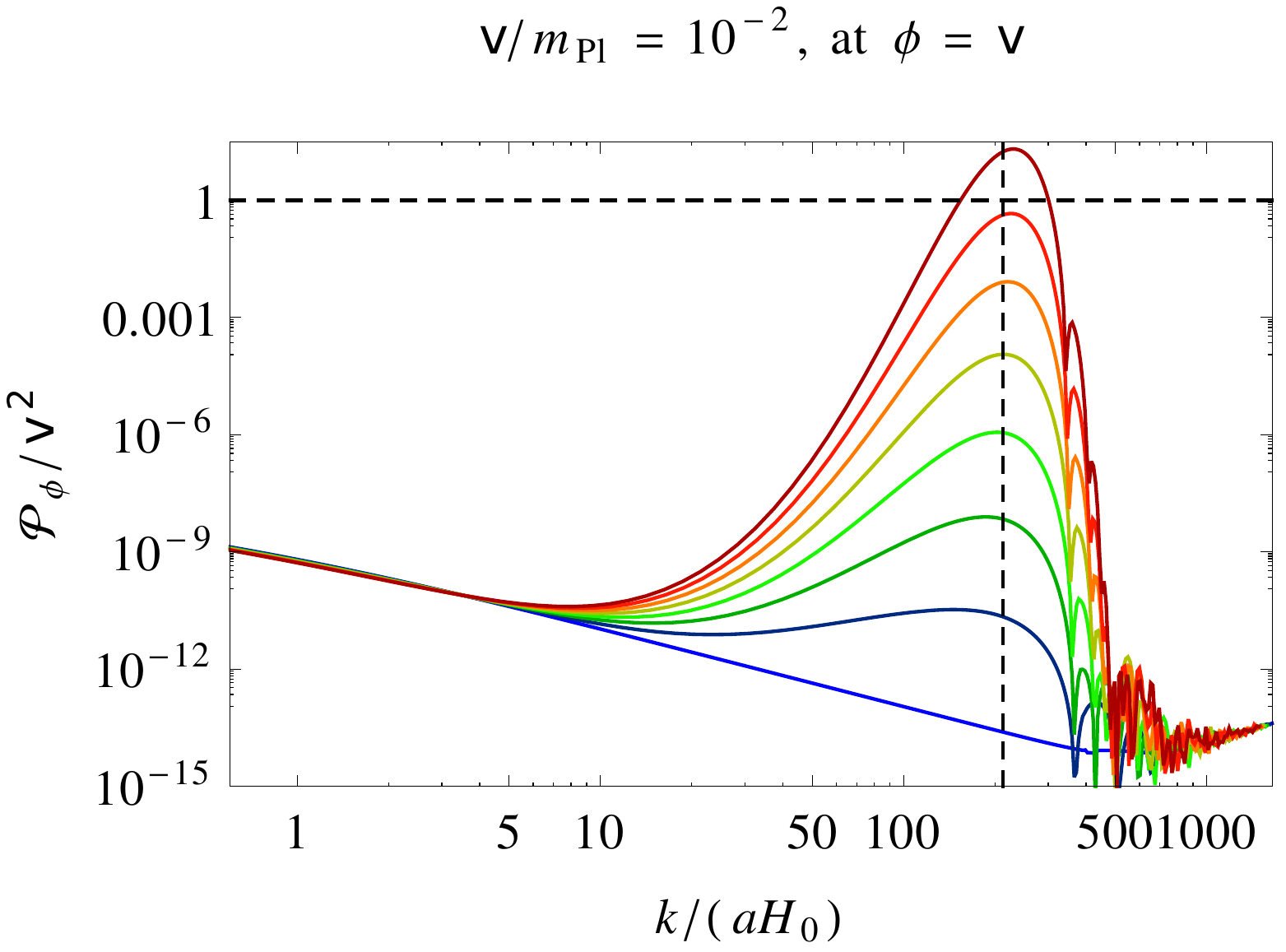} &
\includegraphics[width=0.48\textwidth]{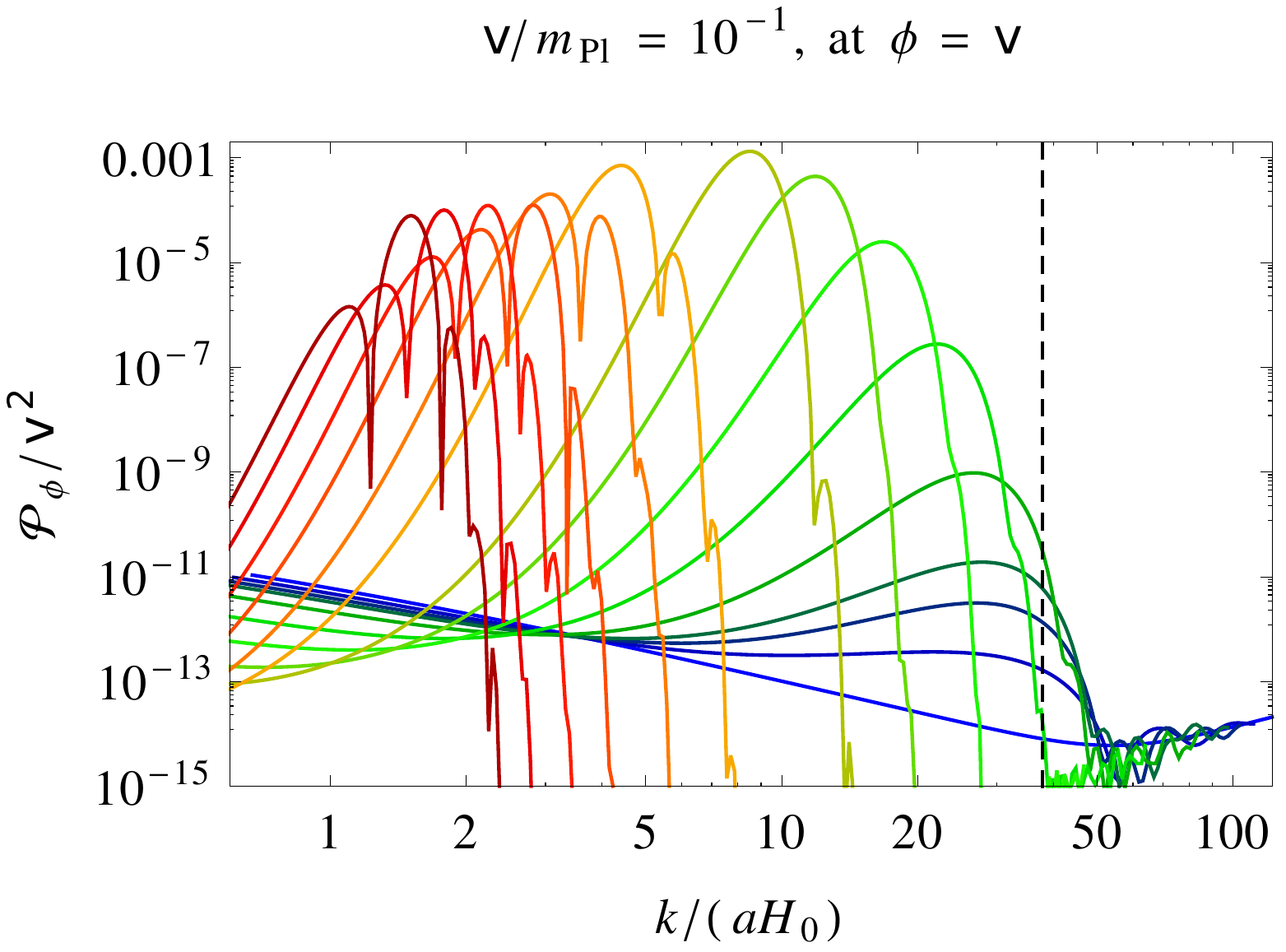}
\end{array}$
  \caption{Linear power spectra evaluated at $\phi=v$ and $\dot{\phi} > 0$ for the first few oscillations (first to last oscillation: blue to red). Tachyonic preheating enhances modes with $k < k_{\rm max} \simeq 0.435 m_{\rm min}$, with the strongest amplification in the infrared $(k \lesssim H_0)$, whereas the oscillations enhance modes with much larger momenta around the peak scale $k_{\rm peak} \simeq 7 (\mpl/v)^{3/4} aH_0$, which is denoted by a vertical dashed line. For $v/\mpl=10^{-1}$, we do not show every oscillation, because linear preheating lasts for very many oscillations; the dark red line in this case corresponds to over 1000 oscillations.}
  \label{fig:linearSpectra}
\end{figure}

\begin{figure}[ptb]
  \centering$
\begin{array}{cc}
\includegraphics[width=0.48\textwidth]{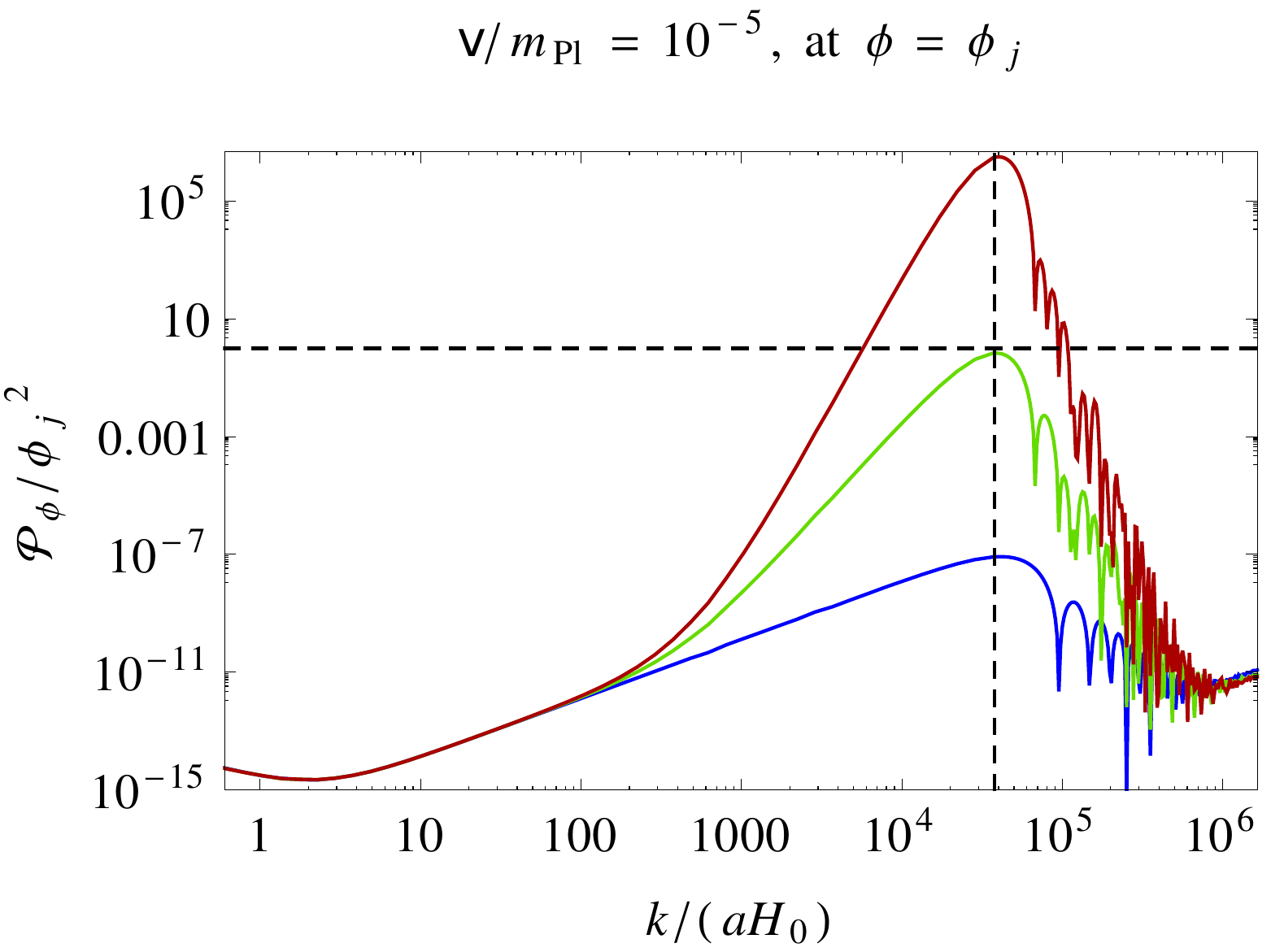} &
\includegraphics[width=0.48\textwidth]{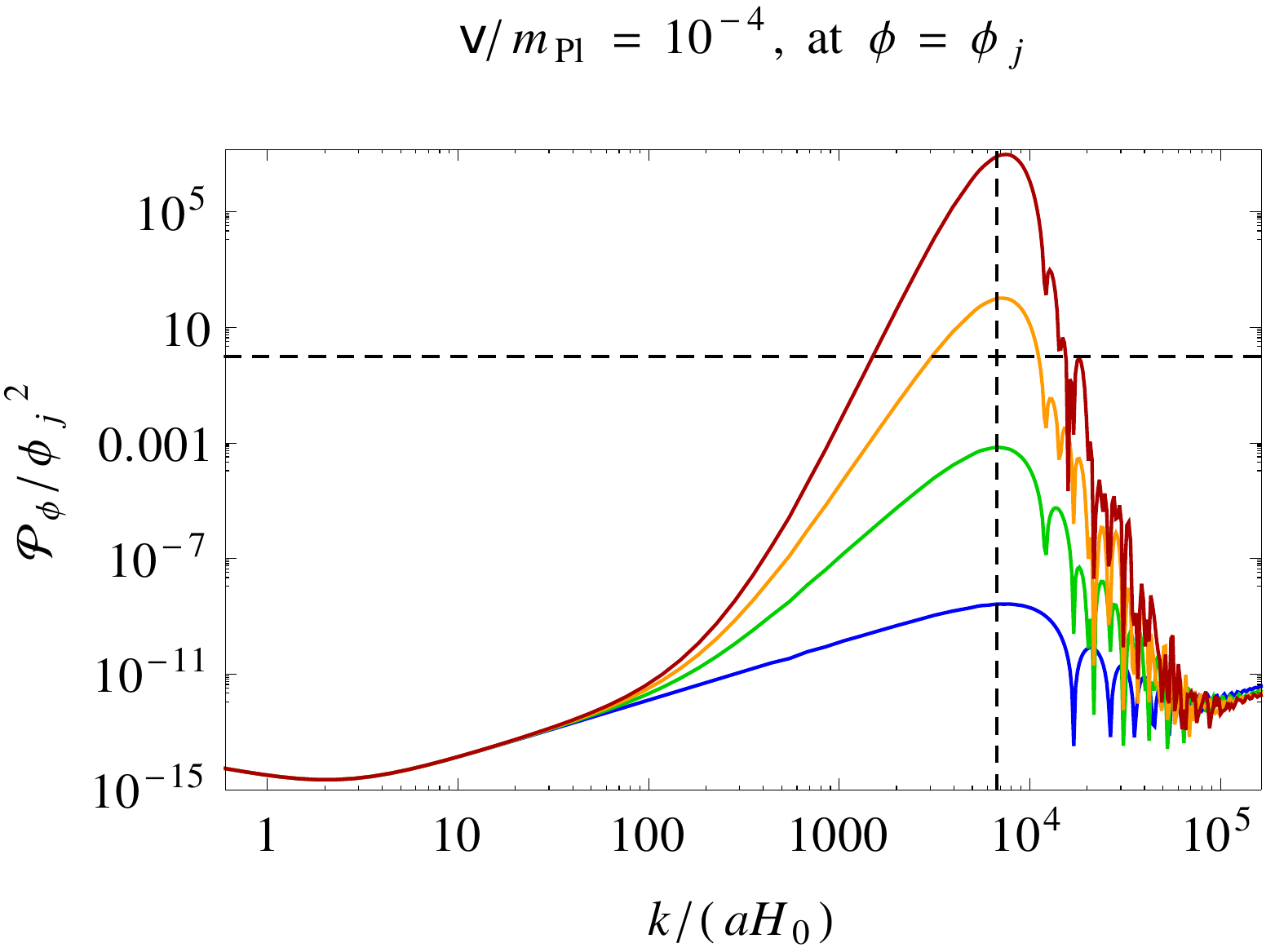} \\
\includegraphics[width=0.48\textwidth]{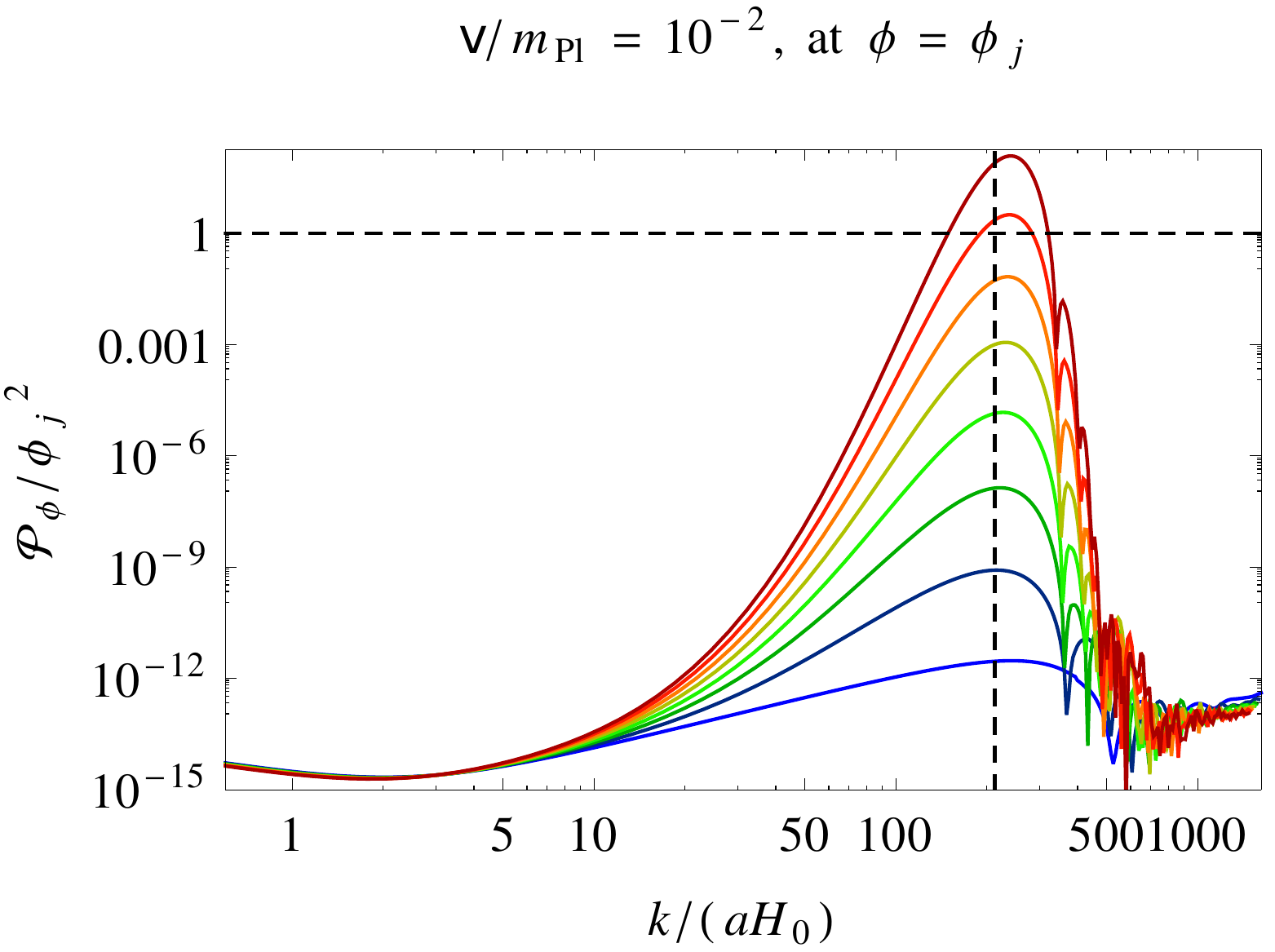} &
\includegraphics[width=0.48\textwidth]{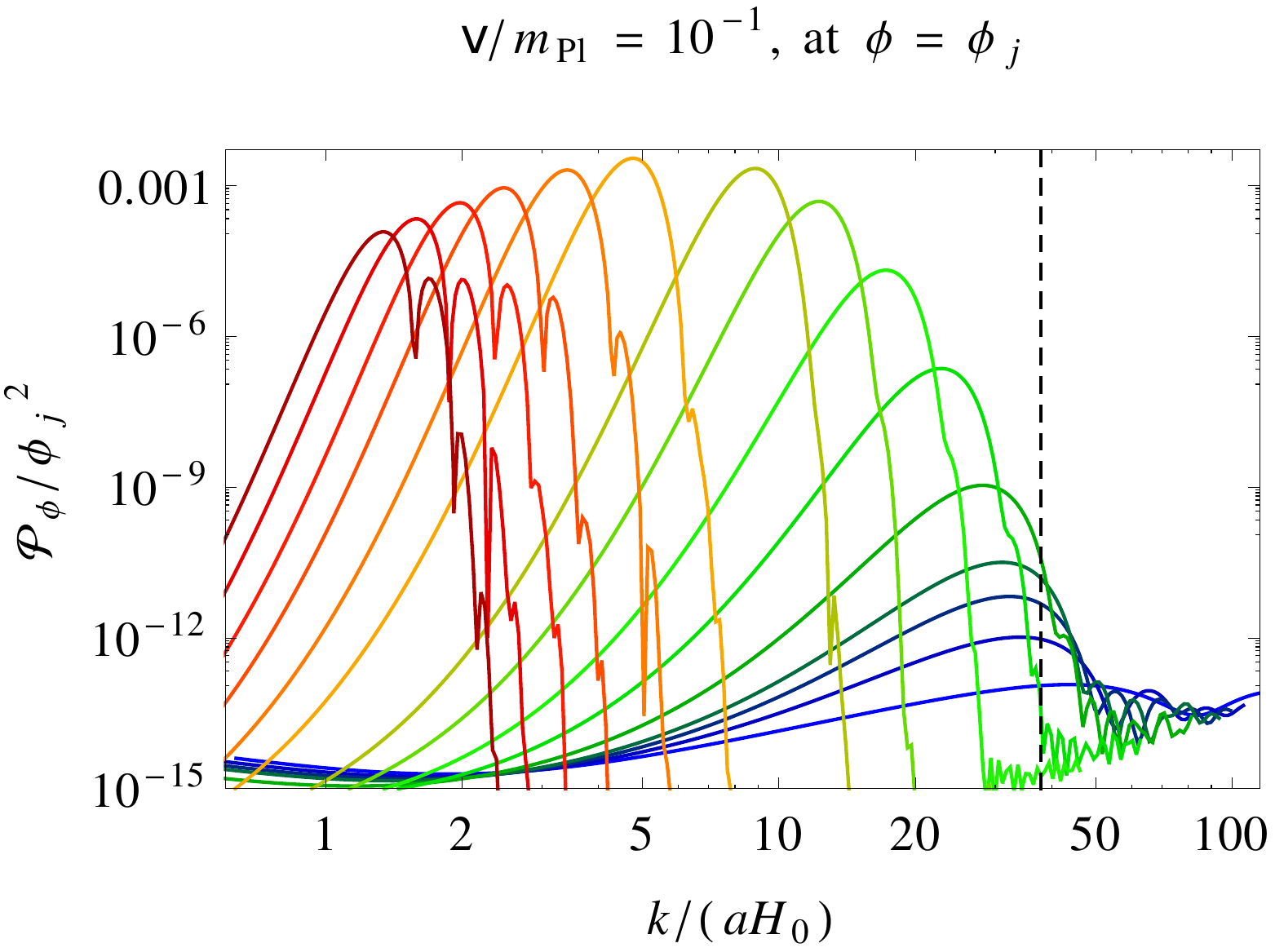}
\end{array}$
  \caption{Linear power spectra normalized to the mean inflaton value $\phi_j$, evaluated at the end of each oscillation when the inflaton is closest to the hilltop (first to last oscillation: blue to red). We see that the infrared peak from tachyonic preheating is negligible near the hilltop, whereas the perturbations around the peak scale $k_{\rm peak}$ (vertical dashed line) can grow amplitudes larger than $\phi_j$, indicating that on these scales, the perturbations can push the inflaton towards $\phi(x) < 0$. For $v/\mpl=10^{-1}$, we do not show every oscillation, because linear preheating lasts for very many oscillations; the dark red line in this case corresponds to over 1000 oscillations.}
  \label{fig:linearSpectraWallCrossing}
\end{figure}

Integrating the mode equations \eqref{eq:eomDeltaPhiK} numerically, we can calculate the power spectrum $\mathcal{P}_\phi(k)$ of the inflaton perturbations. Fig.~\ref{fig:linearSpectra} shows $\mathcal{P}_\phi/\overline{\phi}\vphantom{\phi}^2$ for various values of $v$ evaluated at the time when the inflaton crosses the minimum ($\phi=v$ and $\dot{\phi}>0$), while fig.~\ref{fig:linearSpectraWallCrossing} shows the power spectra closest to the hilltop at the end of each full oscillation, where $\dot{\phi}=0$.

All of the plots show the strong growth of power for modes with $k \sim k_{\rm peak}$ with each successive oscillation. For $v/\mpl \leq 10^{-2}$, it only takes a few oscillations for the perturbations to grow non-linear (i.e., $\mathcal{P}_\phi(k_{\rm peak}) \gtrsim \overline{\phi}\vphantom{\phi}^2$). Only for large $v/\mpl=10^{-1}$, the perturbations remain small due to Hubble damping, as explained above.

For the infrared modes, fig.~\ref{fig:linearSpectra} clearly shows the tachyonic preheating spectrum with $\mathcal{P}_\phi(aH_0) \sim 10^{-13}\mpl^2$ and $\mathcal{P}_\phi(k) \propto k^{-2}$. The infrared modes do not continue to grow during the tachyonic oscillation phase: for $v/\mpl \leq 10^{-2}$, the spectra after the $j$-th oscillation lie on top of each other. For $v/\mpl = 10^{-1}$, we even see that the infrared modes slowly decay due to Hubble damping and redshifting. Near the hilltop, the tachyonic preheating spectrum almost vanishes, as one can see in fig.~\ref{fig:linearSpectraWallCrossing}. The infrared part of the spectrum actually oscillates during the tachyonic oscillation phase, growing to maximum values at $\phi \simeq v$ and shrinking to minimal values at $\phi \simeq \phi_j$, whereas $\mathcal{P}_\phi(k_{\rm peak})/\overline{\phi}\vphantom{\phi}^2$ continuously grows throughout each oscillation\footnote{To be precise, $\mathcal{P}_\phi(k_{\rm peak})/\overline{\phi}\vphantom{\phi}^2$ temporarily drops when $\phi > v$, but it grows both on its way up and down the hilltop for $\phi < v$.} until the evolution becomes non-linear.

\subsection{End of linear preheating and hill crossing}

Linear preheating predicts that for $10^{-5} \lesssim v/\mpl \lesssim 10^{-2}$, there is a rapid growth of perturbations at the scale $k_{\rm peak}$, producing a large fluctuation amplitude
\begin{align}
 \braket{\delta \phi^2(\vec{x})} \, = \, \int(d \ln k) \mathcal{P}_\phi(k) \, \sim \, \mathcal{P}_\phi(k_{\rm peak}) \, \gtrsim \, \overline{\phi}\vphantom{\phi}^2.
\end{align}
This could indicate either that the fluctuations become large enough to push the inflaton field over the hilltop towards $\phi(\vec{x}) < 0$, or that linear perturbation theory breaks down and non-linear interactions stop the growth of perturbations before this happens. To find out which of these options is realized, we need to go beyond the linear approximation.

As we will show below, it turns out that the fluctuations can really push $\phi(\vec{x})$ to negative values at some points in space, and the potential subsequently drives $\phi(\vec{x}) \rightarrow -v$ in these regions. The inflaton field is then no longer well described by a homogeneous field $\overline{\phi}$ plus small perturbations $\delta \phi(\vec{x})$. To study the formation and evolution of the regions with $\phi(\vec{x}) < 0$, we need to solve the field equations of motion non-perturbatively using numerical lattice simulations.

\section{Non-linear dynamics}
\label{sec:lattice}

In this section, we present and discuss the results of numerical lattice simulations of the non-linear stage of preheating for $v/\mpl = 10^{-2}$ and $v/\mpl = 10^{-5}$.

In the previous section, we have seen that for $10^{-5} \lesssim v/\mpl \lesssim 10^{-1}$, the inflaton fluctuations get amplified during the initial descent towards $\phi=v$ at scales $k_{\rm tac}\sim aH_0$. Then the system oscillates between $\phi \ll v$ and $\phi > v$ during which time the spectrum gets amplified at scales $k_{\rm peak}\sim (\mpl/v)^{3/4}aH_0$, either until non-linearities become dominant (for $v/\mpl \lesssim 10^{-2}$) or until the oscillations become damped and redshifted by the Hubble expansion (for $v/\mpl \gtrsim 10^{-1}$).

In order to study the non-linear phase of preheating, we numerically solve the non-linear equations of motion
\begin{align}
&\ddot{\phi}(t,\bar{x}) + 3H\dot{\phi}(t,\bar{x}) - \frac{1}{a^2}\bar{\nabla}^2\phi(t,\bar{x}) + \frac{\partial V}{\partial \phi} \, = \, 0\,, \label{eq:nleom1}\\ 
&H^2\, =\,\frac{1}{3\mpl^2}\left\langle V  +  \frac{1}{2}\dot{\phi}^2 + \frac{1}{2a^2} \left|\bar{\nabla}\phi\right|^2 \right\rangle\,, \label{eq:nleom2}
\end{align}
where $\bar{\nabla}$ are gradients with respect to the comoving coordinates $\bar{x}$. We use a modified version of the program LATTICEEASY \cite{Felder:2000hq} to solve these equations on a discrete spacetime lattice.

The hierarchy between $k_{\rm tac}$ and $k_{\rm peak}$ implies that the lattice must have much more than $(\mpl/v)^{3/4}$ points per dimension to include both scales in the simulation. For small $v/\mpl \sim 10^{-5}$, this becomes impractical due to our limited computing power.

However, from the linear analysis we expect the fluctuations at $k_{\rm tac}$ to remain relatively small, as only the fluctuations at the peak scale $k_{\rm peak}$ grow during the oscillations in the linear phase. At the end of the linear phase, most of the perturbations will be around $k \sim k_{\rm peak}$. We therefore ignore the large wavelength modes at $k_{\rm tac}$ and only include scales around $k_{\rm peak}$ in order to resolve the dominant small distance fluctuations.

For $v/\mpl = 10^{-2}$, we could even keep most of the infrared modes, because the hierarchy between $k_{\rm tac}$ and $k_{\rm peak}$ is smaller. However, the linear power spectrum at $k_{\rm tac}$ is very small for $v/\mpl = 10^{-2}$, so even in this case we prefer to cut off some infrared power in favour of a larger UV cutoff, which is more important to resolve the small-distance fluctuations produced by the hill crossing.

To resolve a reasonable range around $k_{\rm peak}$, we focus on $2$ spatial dimensions, although we also did some lower-resolution runs in $3$ dimensions (see section~\ref{sec:3D}) and saw the same qualitative behaviour.

\subsection{Initialization of the lattice}
\label{sec:latticeInit}

 The initial time of the lattice simulation must be chosen such that the system is still well described by the linear equations~\eqref{eq:eomDeltaPhiK}.
 We choose to start after $\eta=-1$ but shortly before $\varepsilon=\frac12\mpl^2(V'/V)^2=1$, with initial conditions and parameters shown in table~\ref{tab:ini}.
 
\begin{figure}[ptb]
  \centering$
\begin{array}{cc}
\includegraphics[height=5cm]{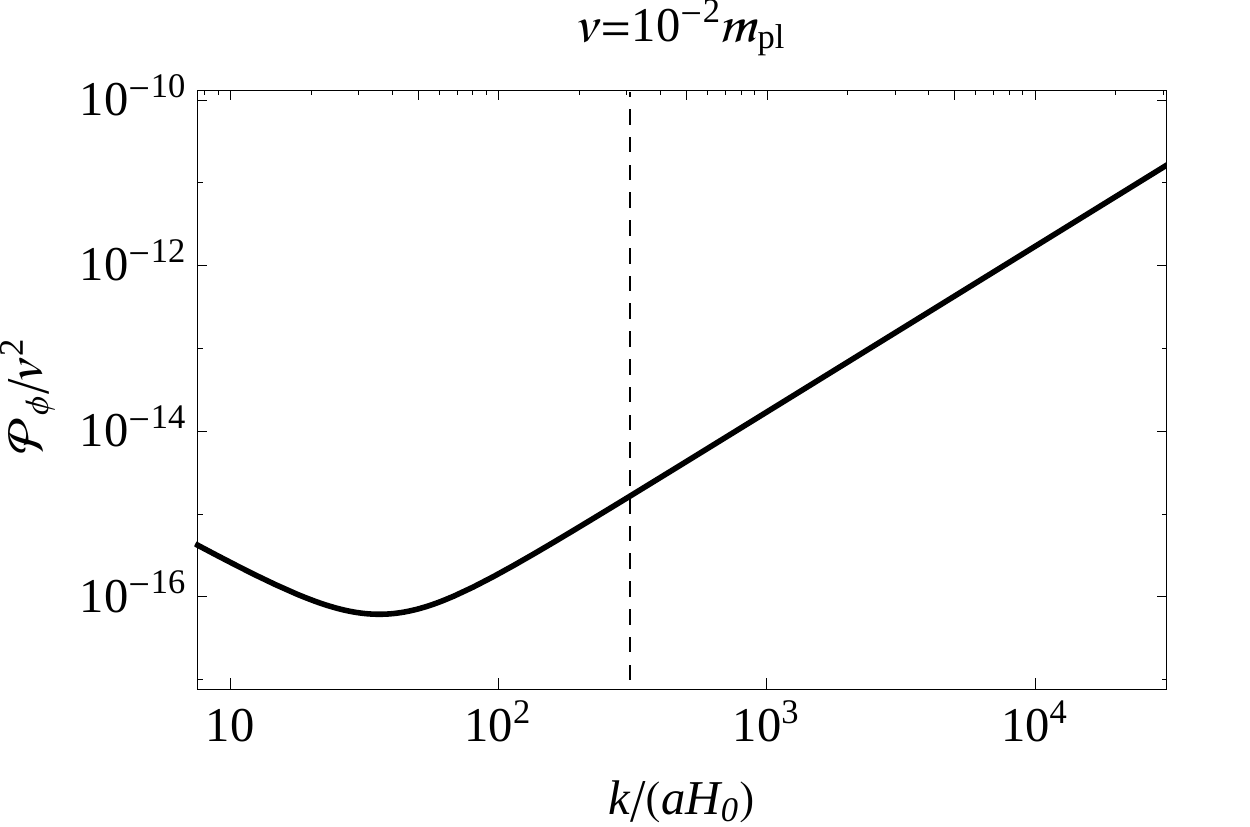}
\includegraphics[height=5cm]{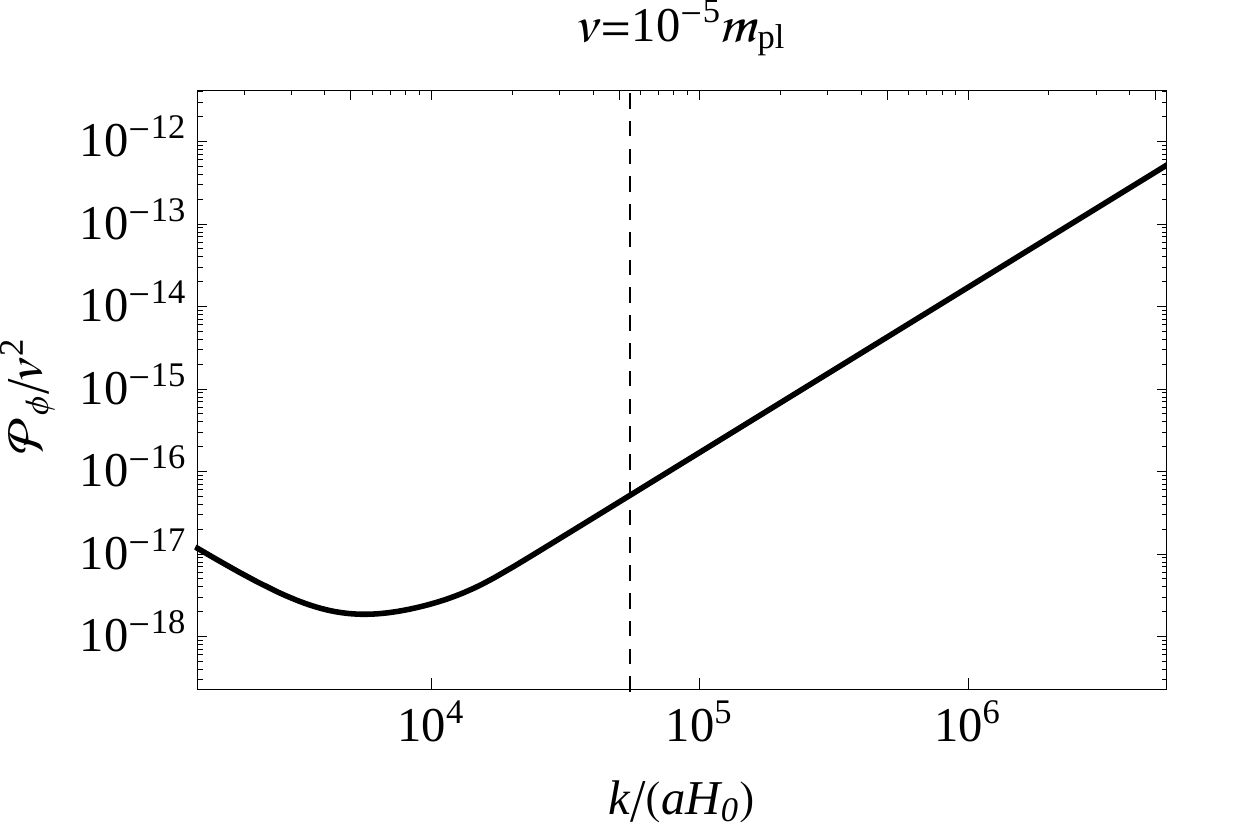}
\end{array}$
  \caption{Initial spectra $\mathcal{P}_{\phi}/v^2$ of our lattice simulations for $v=10^{-2}\mpl$ (left) and $v=10^{-5}\mpl$ (right), with other parameters given in table~\ref{tab:ini}. The vertical dashed line corresponds to $k=k_{\rm peak}$ and the range of scales goes from $k_{\rm ir} = k_{\rm uv}/4096$ to $k_{\rm uv} = 100k_{\rm peak}$. For $k\gtrsim 0.1k_{\rm peak}$, the spectra are dominated by vacuum fluctuations $\mathcal{P}_{\rm vac} \propto k^2$, whereas for $k\lesssim 0.1k_{\rm peak}$ one can see the effect of tachyonic preheating. The scale $k_{\rm tac}$ is outside of the range for both  $v=10^{-2}\mpl$ and $v=10^{-5}\mpl$, which we expect not to affect our result significantly as perturbations on that scale are subdominant compared to fluctuations at $k_{\rm peak}$ (see section~\ref{sec:linear}).}
  \label{fig:spectrain}
\end{figure}

 The initial field fluctuations $\langle\phi^2_k\rangle_{\rm lin}$ for $v=10^{-2}\mpl$ and $v=10^{-5}\mpl$ are extracted from the linear equations~\eqref{eq:eomDeltaPhiK}. As shown in fig.~\ref{fig:spectrain}, the power spectra  around $k_{\rm peak}$ at that time are mainly dominated by the vacuum contribution. The tachyonic contribution becomes important for scales $k\lesssim 0.1k_{\rm peak}$, corresponding to the infrared tail of our initial spectra.

As usual, the norm of the fluctuations is initialized as a stochastic variable obeying the probability distribution \cite{Polarski:1995jg,Khlebnikov:1996mc}
\be
PD(|\phi_k|)=\frac{2|\phi_k|}{\langle\phi^2_k\rangle_{\rm lin}}\exp\left[-\frac{|\phi_k|^2}{\langle\phi^2_k\rangle_{\rm lin}}\right]\,.
\label{eq:pdphi}
\en
The initial Fourier components of $\phi$ and their derivatives are given by
\be
\phi_k &=& \frac{1}{a}\frac{|\phi_k|}{\sqrt{\alpha_{+}^2+\alpha_{-}^2}}\left(\alpha_{+}e^{i2\pi\theta_+ +i k t} + \alpha_{-}e^{i2\pi\theta_- -i k t}\right)\,,\nn
\dot{\phi}_k &=& \frac{i k}{a}\frac{|\phi_k|}{\sqrt{\alpha_{+}^2+\alpha_{-}^2}}\left(\alpha_{+}e^{i2\pi\theta_+ +i k t} - \alpha_{-}e^{i2\pi\theta_- -i k t}\right) -H\phi_k\,.
\label{eq:phifl}
\en
where $\alpha_+$, $\alpha_-$, $\theta_+$ and $\theta_-$ are random real numbers uniformly distributed between $0$ and $1$.\footnote{Note that this is different from the original LATTICEEASY initialization, where the fluctuations are initialized as standing waves, corresponding to $\alpha_-=\alpha_+=1$.}

In setting the derivative $\dot{\phi_k}$, we assumed that $\phi_k\propto e^{\pm ikt}$. Although this is true only for the vacuum fluctuations and not for the infrared part $k\lesssim 0.1k_{\rm peak}$ of the the spectra in fig.~\ref{fig:spectrain}, our results are not affected by this assumption. Indeed, the lattice simulations' results during the linear phase match those of the linear analysis.

The fluctuations \eqref{eq:phifl} are then Fourier transformed to position space and a discretized version of the equations of motion is solved on a lattice with spacing $\delta x = 2\pi/k_{\rm uv}$, with $k_{\rm uv}=100k_{\rm peak}$ as the ultra-violet cutoff. The size of the lattice is given by $L=2\pi N/ k_{\rm uv}$ where $N$ is the number of points per dimension. This defines the infrared cutoff $k_{\rm ir} = k_{\rm uv}/N$. In section~\ref{sec:2dsim} we present results of simulations in $D=2$ spatial dimensions with $N=4096$ and in section~\ref{sec:3D} we briefly discuss simulations in $D=3$ spatial dimensions with $N=256$. 

In order to run the simulations, we must choose suitable rescalings. We define our program units to be
\be
\phi_{\rm pr} = a^{3/2}\phi\,,\;\;\;\;\;\bar{x}_{\rm pr}= k_{\rm peak}\bar{x}\,,\;\;\;\;\;dt_{\rm pr}=k_{\rm peak}dt\,,
\label{eq:LEres}
\en
This corresponds to choosing LATTICEEASY rescalings $A=1$, $B=k_{\rm peak}$, $r=3/2$ and $s=0$.

\begin{table}[t]
\centering
\begin{tabular}[c]{ | c || c | c |c | c| c|}
\hline
$v/\mpl$ & $\langle\phi\rangle_{\rm i}/v$ & \vphantom{$\frac{\dot{f}}{\dot{f}}$}$\langle\dot{\phi}\rangle_{\rm i}/v^2$ & $k_{\rm uv}$ & $N$ & $LH_0$\\
\hline
\vphantom{$\frac{f}{f}$} $10^{-2}$ & $0.058$ & $8.9\times10^{-10}$ & $100k_{\rm peak}$ & $4096$ &  $0.83$\\ 
\hline
\vphantom{$\frac{f}{f}$} $10^{-5}$ & $0.0098$ & $2.7\times10^{-11}$ & $100k_{\rm peak}$ & $4096$ & $4.7\times10^{-3}$ \\
\hline
\end{tabular}
\caption{ Initial field and derivative mean values and other parameters of the 2D lattice simulations.}
\label{tab:ini}
\end{table}

\subsection{Results of lattice simulations in $2$ dimensions}
\label{sec:2dsim}

Based on the linear analysis of section~\ref{sec:linearPowerSpectra}, we expect that for $v=10^{-1}\mpl$ the dynamics remain linear throughout the oscillations phase. Our lattice simulations confirm this result, with the variance growing to a maximal value of $\langle\delta\phi^2\rangle\simeq10^{-5}v^2$.

For $v \lesssim 10^{-6}\mpl$, on the other hand, preheating is expected to become non-linear already during the tachyonic preheating stage \cite{Brax:2010ai}, which should prevent a tachyonic oscillations phase, so that no spectral peak at $k_{\rm peak}$ is expected.

For these reasons, we focus on the lattice simulation results for $v/\mpl = 10^{-2}$ and $v/\mpl = 10^{-5}$. In $2$ spatial dimensions we can include $N=4096$ points per dimension, which is sufficient to resolve the relevant scales around $k_{\rm peak}$ for both of our choices of $v$.

\subsubsection{Position space slices}
\label{sec:latticeSlices}

\begin{figure}[hp]
  \centering
\text{$v=10^{-2}\mpl$,$\;\;$ $N=1024$,$\;\;$ $D=2$, $\;\;$ $k_{\rm uv} = 100k_{\rm peak}$.}\\$
\begin{array}{cc}
\includegraphics[width=0.33\textwidth]{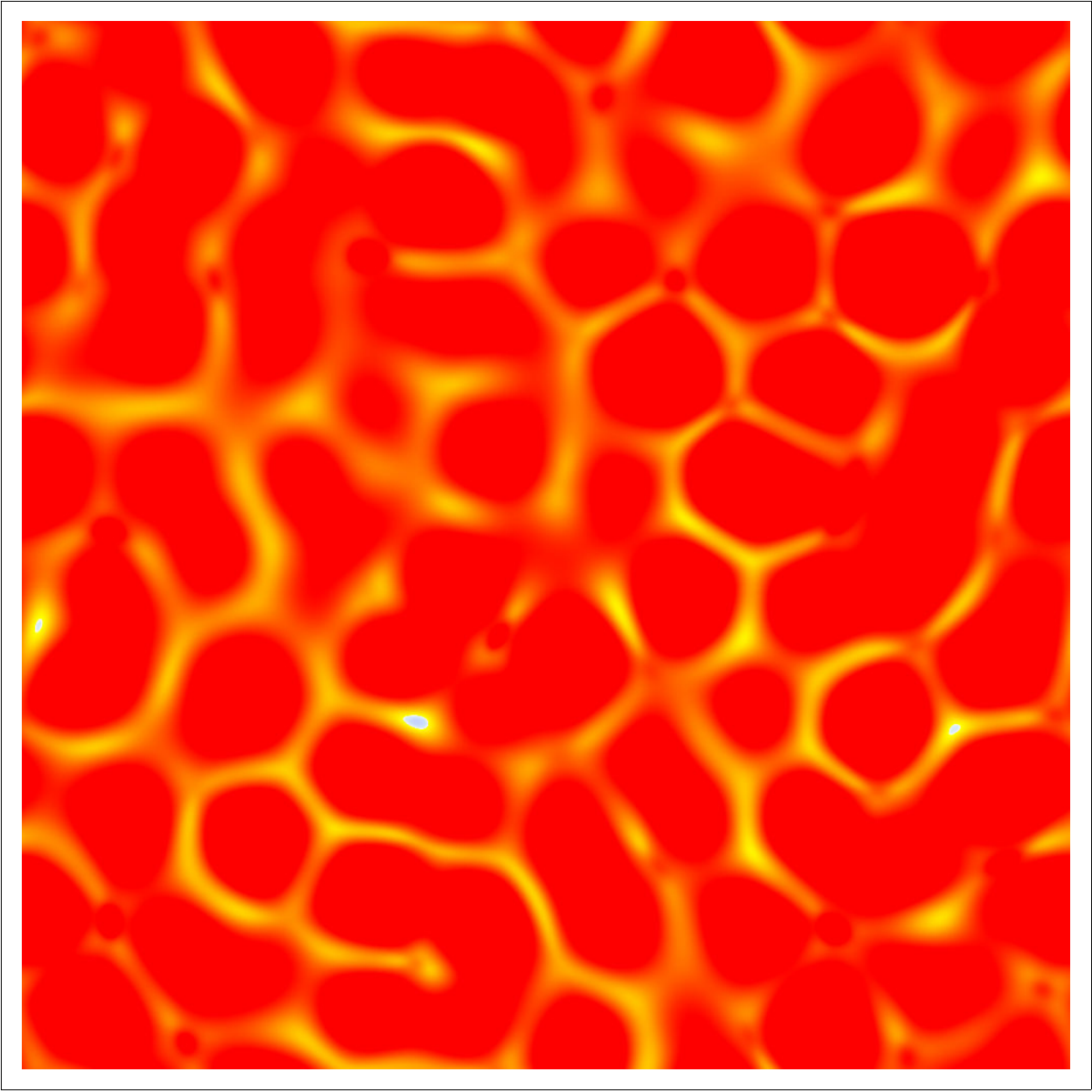} 
\includegraphics[width=0.33\textwidth]{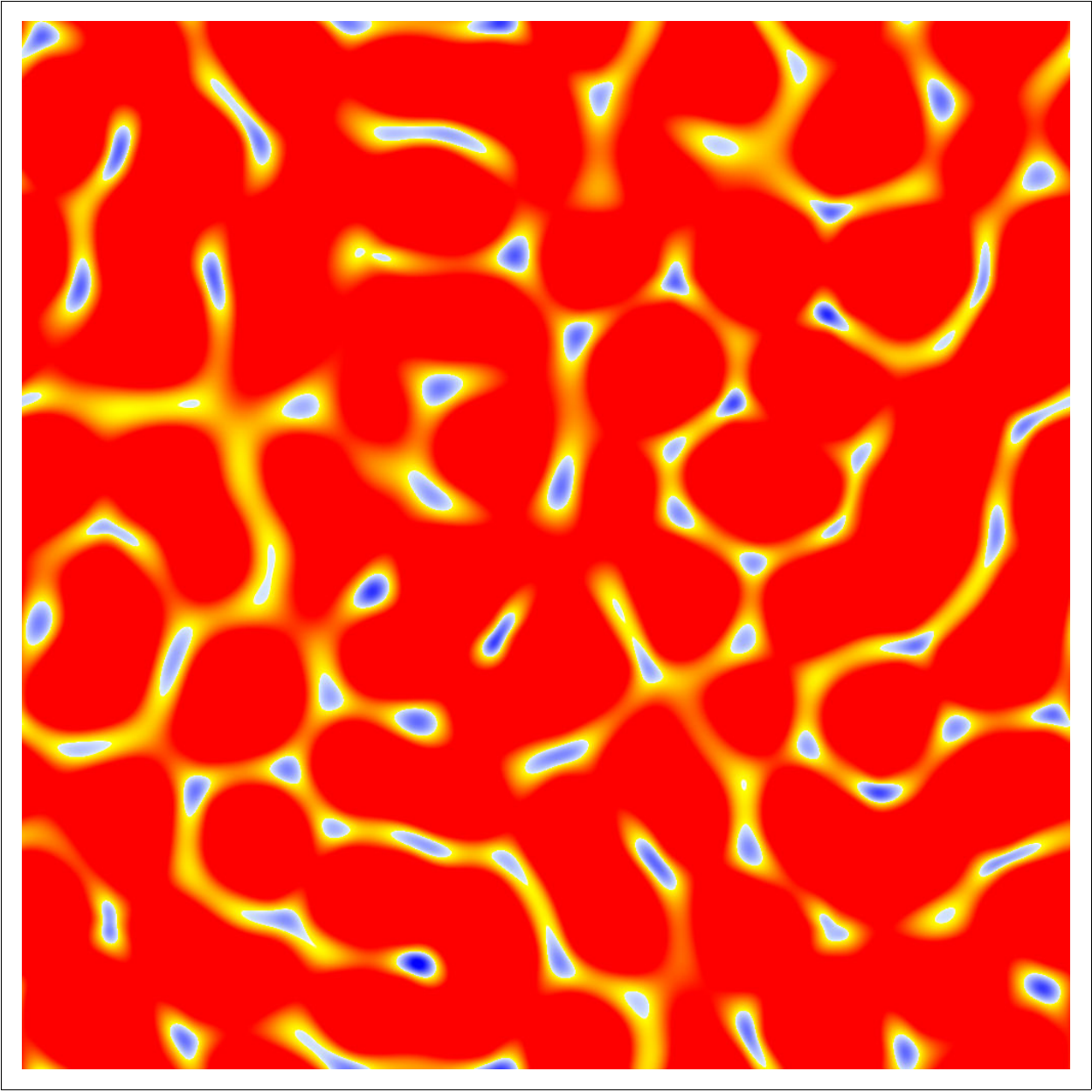} 
\includegraphics[width=0.33\textwidth]{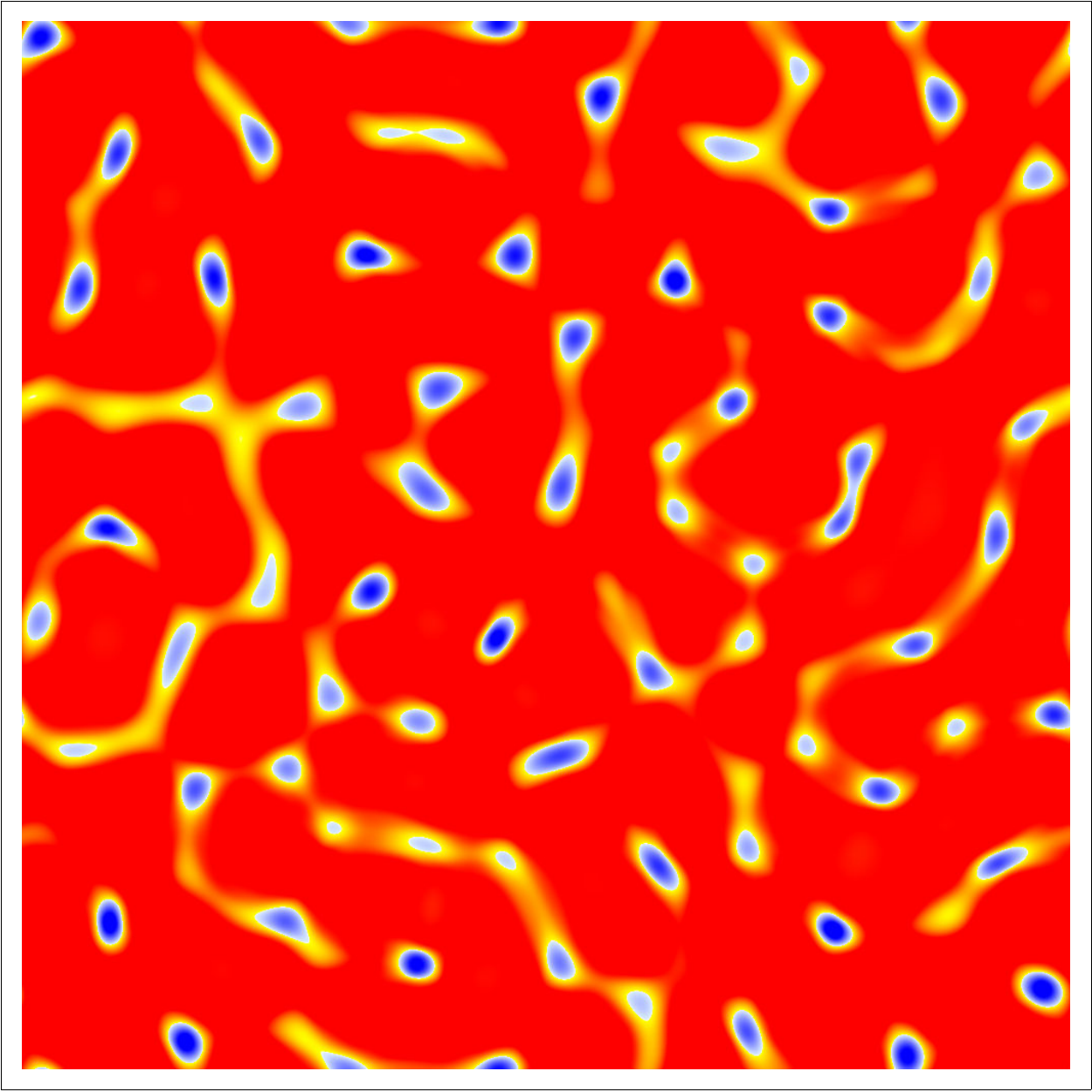}\\
\includegraphics[width=0.33\textwidth]{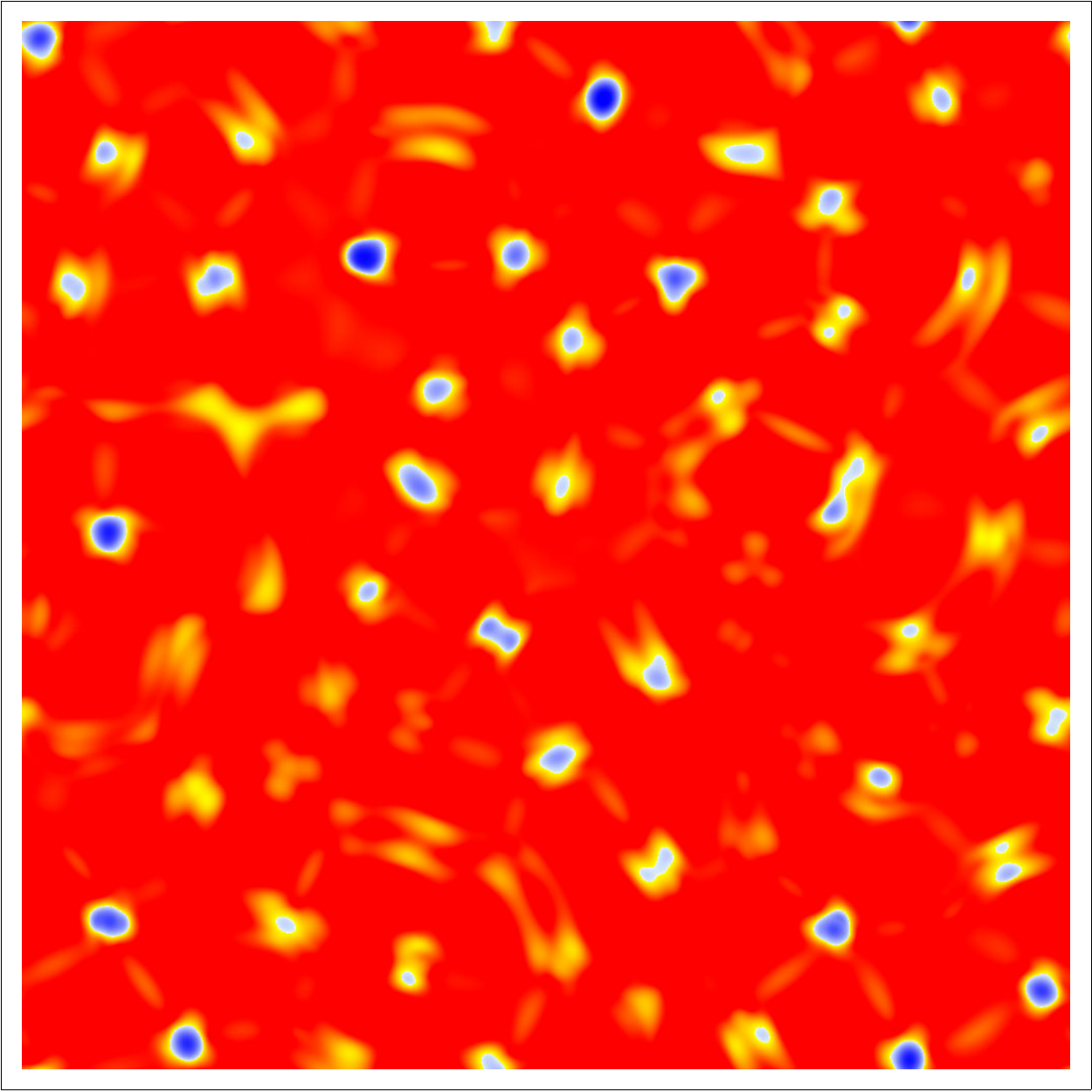}
\includegraphics[width=0.33\textwidth]{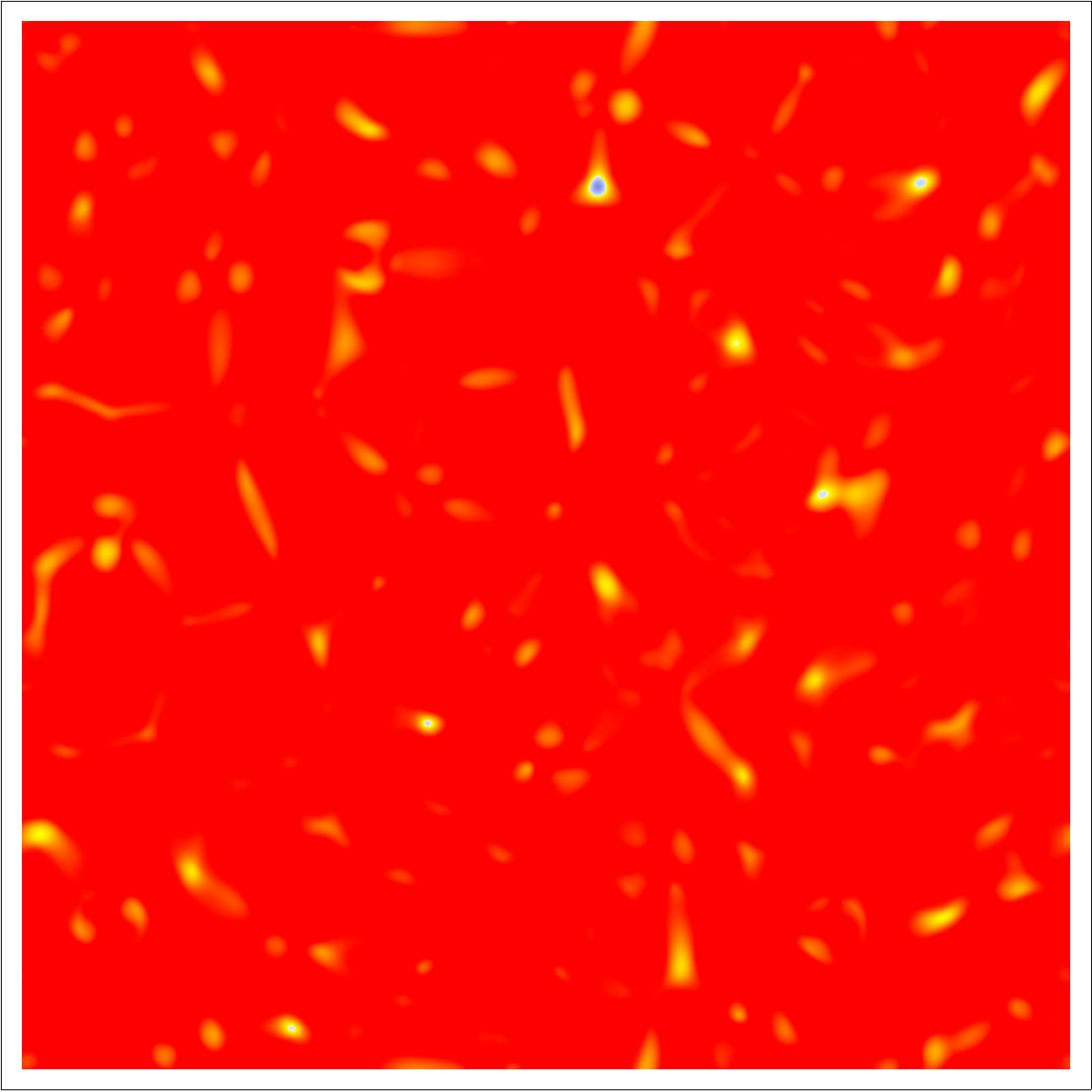} 
\includegraphics[width=0.33\textwidth]{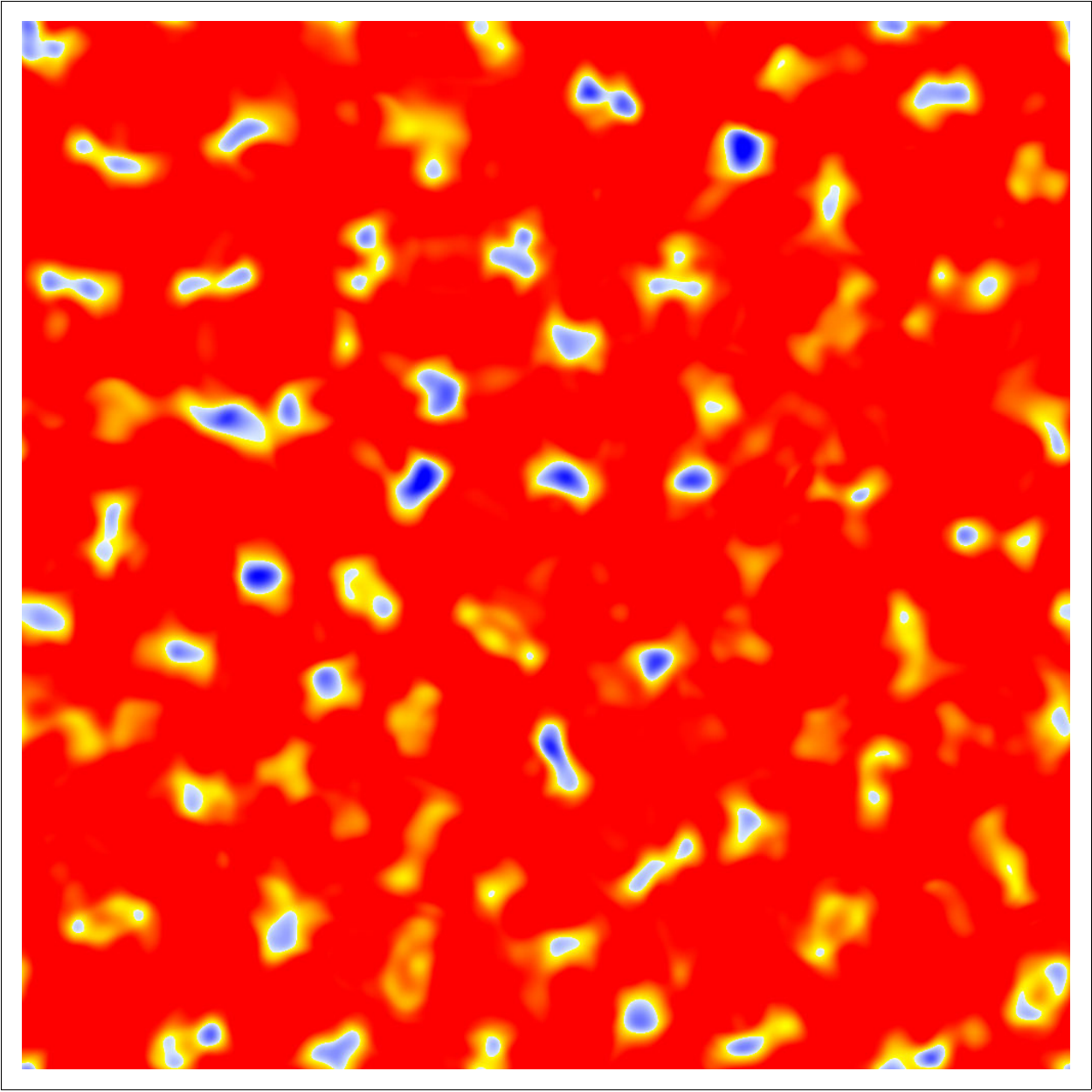}\\ 
\includegraphics[width=0.33\textwidth]{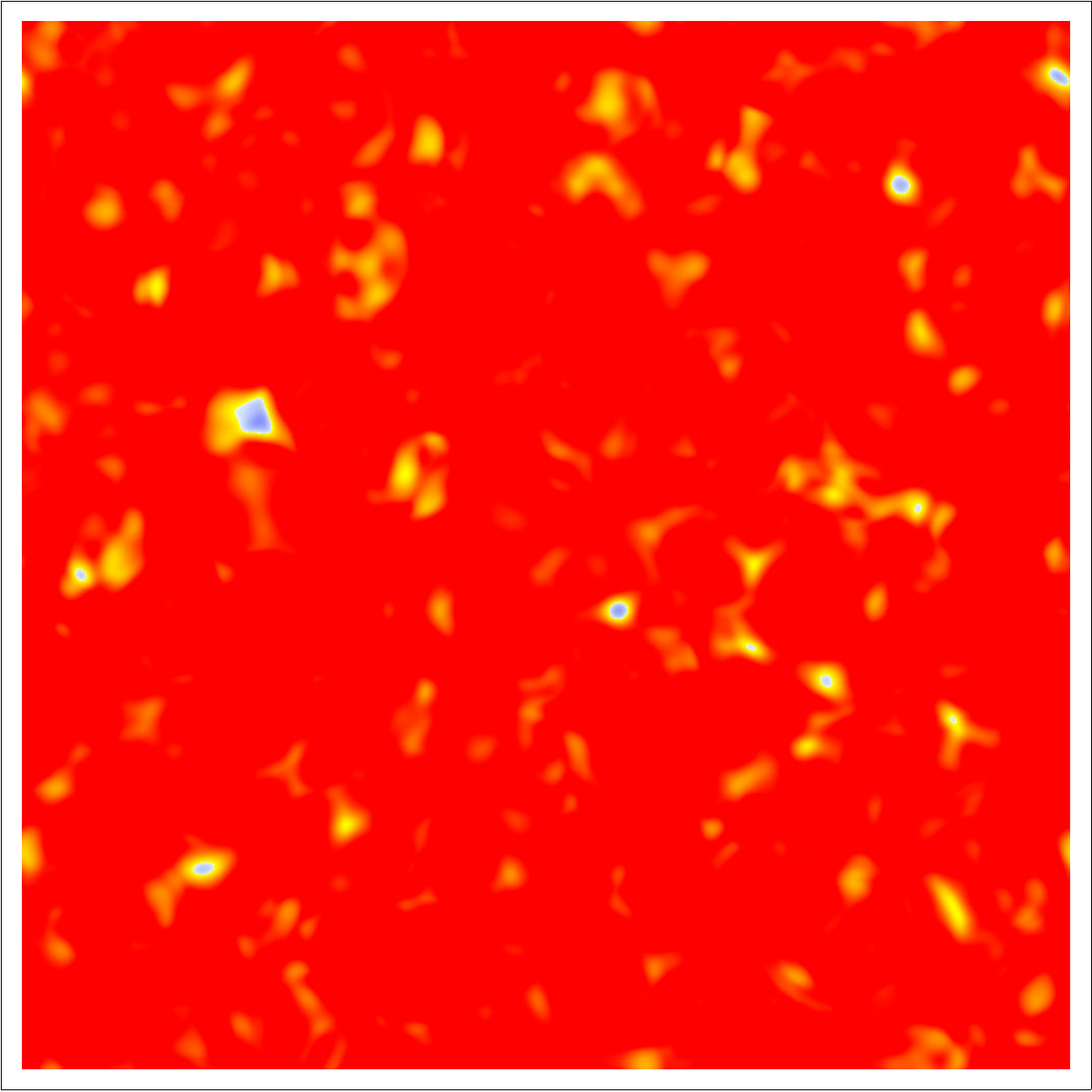}
\includegraphics[width=0.33\textwidth]{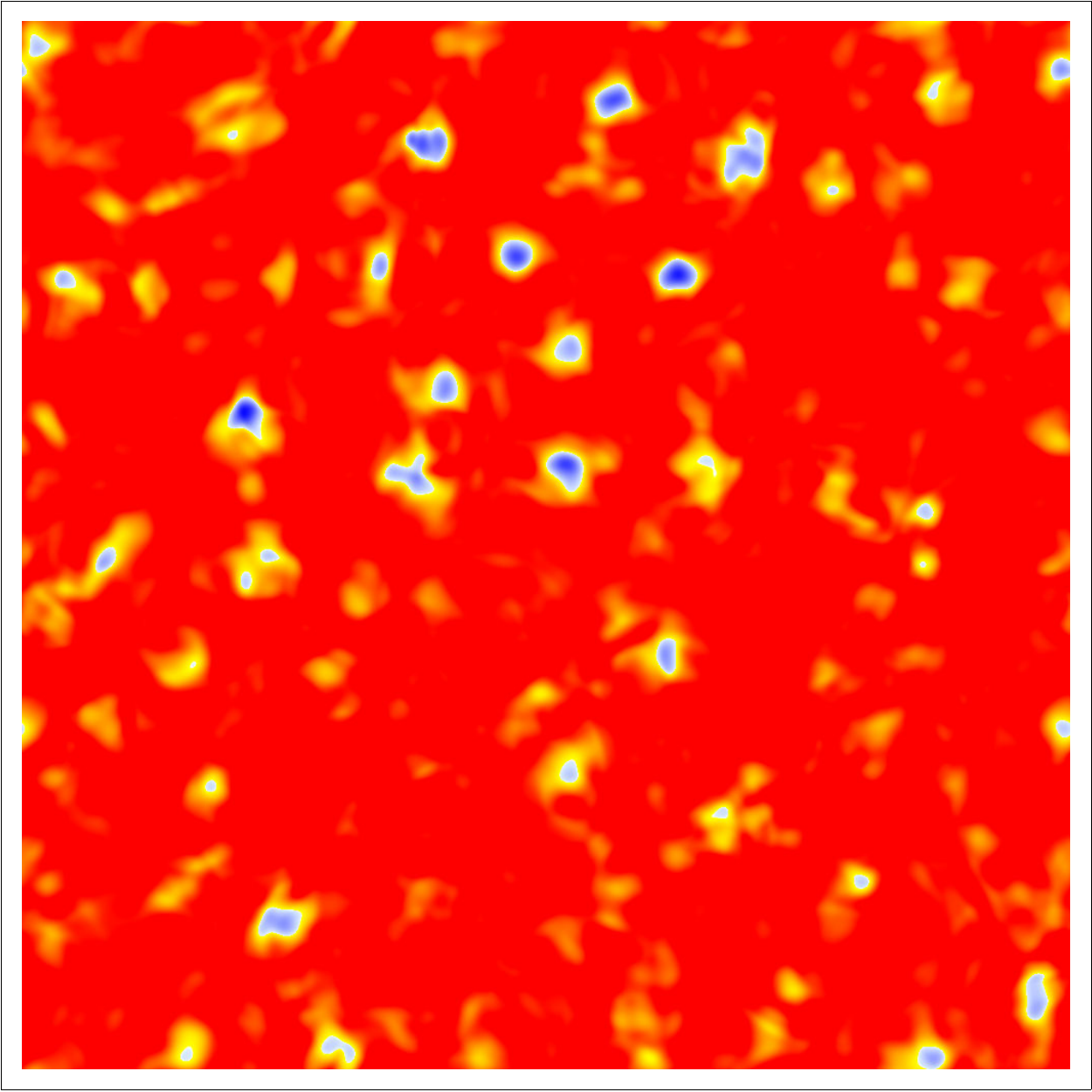} 
\includegraphics[width=0.33\textwidth]{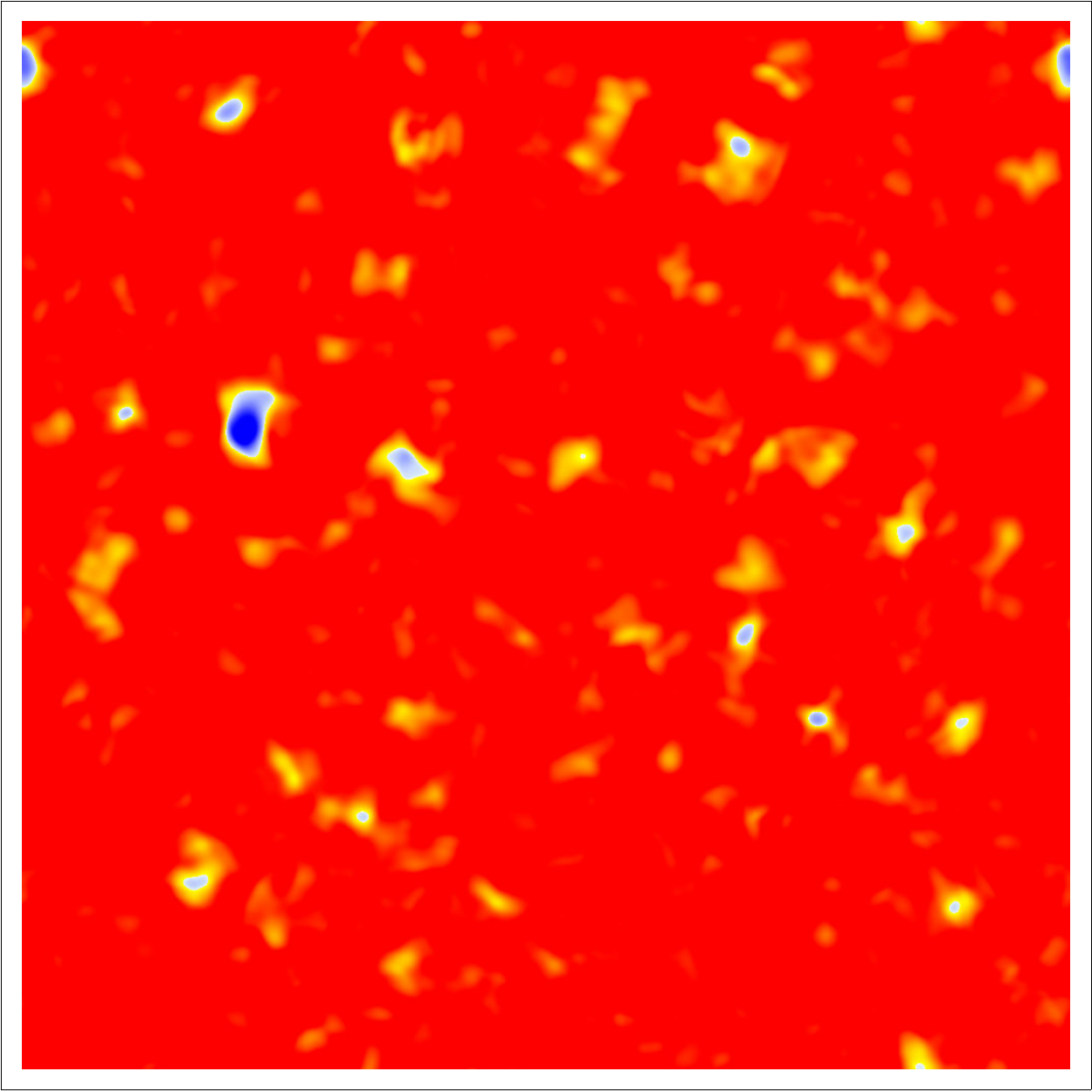}\\ 
\includegraphics[width=0.75\textwidth]{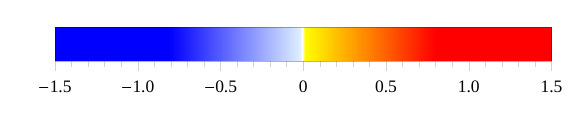}
\end{array}$
  \caption{$\phi(x)$ from the lattice simulation at times $44.25$, $44.75$, $45.25$, $46$, $47.1$, $49.23$, $51.1$, $53.5$, $55.2$ in units of  $k_{\rm peak}^{-1}$. The corresponding ratios of points with $\phi<0$ can be read off fig.~\ref{fig:v2}, where the slices correspond to the blue dots in the upper right plot. The legend is shown in units of $v$. The simulation clearly shows the initial overshooting, formation and dissipation of a filament structure, and eventually localized bubbles oscillating in phase between $\phi \sim -v$ and $\phi \sim v$, as discussed in section~\ref{sec:latticeSlices}. A higher time resolution movie of the evolution can be downloaded at \cite{baselAnimation}.}
  \label{fig:v2slices}
\end{figure}

\begin{figure}[hp]
  \centering
\text{$v=10^{-5}\mpl$,$\;\;$ $N=1024$,$\;\;$ $D=2$, $\;\;$ $k_{\rm uv}=100k_{\rm peak}$}\\$
\begin{array}{cc}
\includegraphics[width=0.33\textwidth]{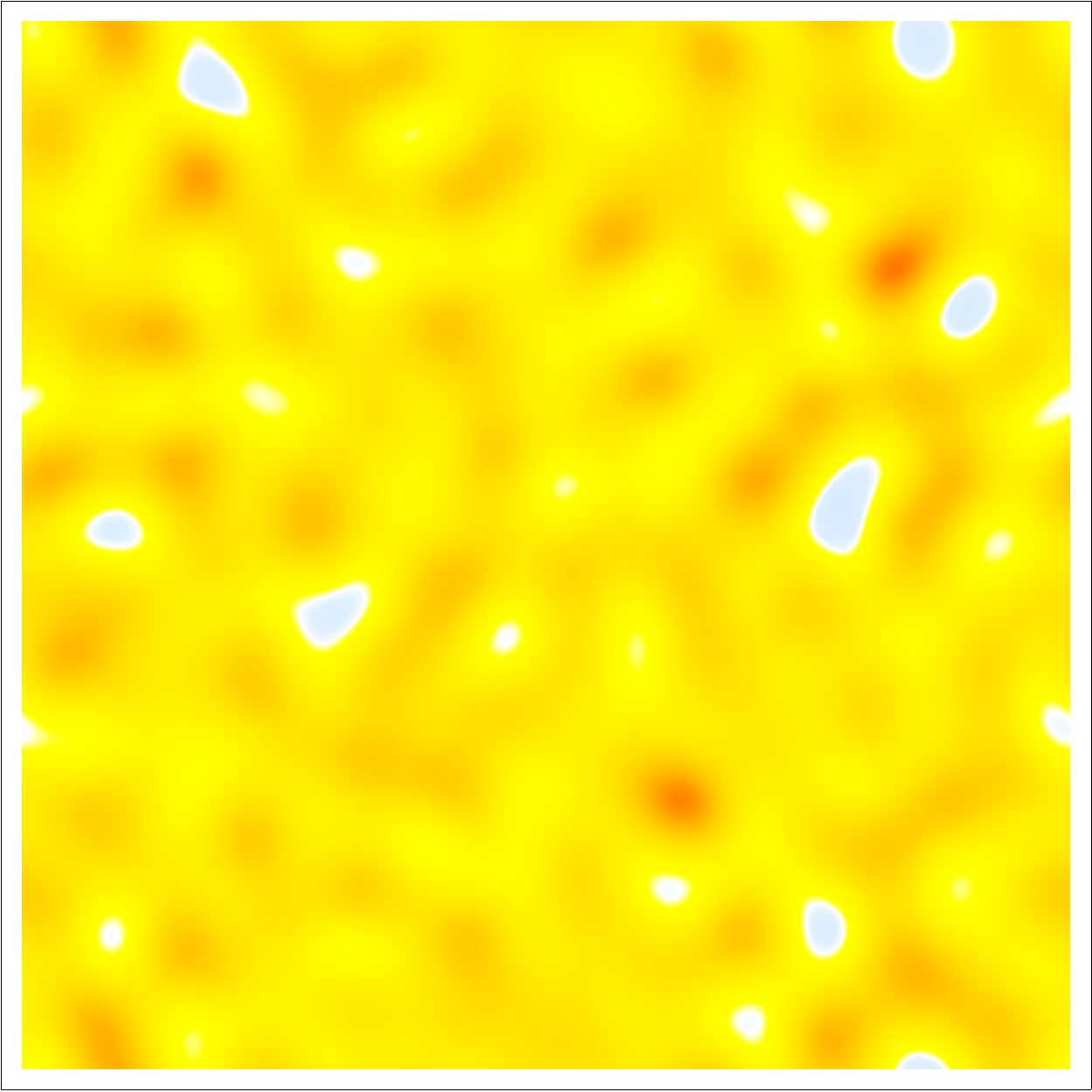} 
\includegraphics[width=0.33\textwidth]{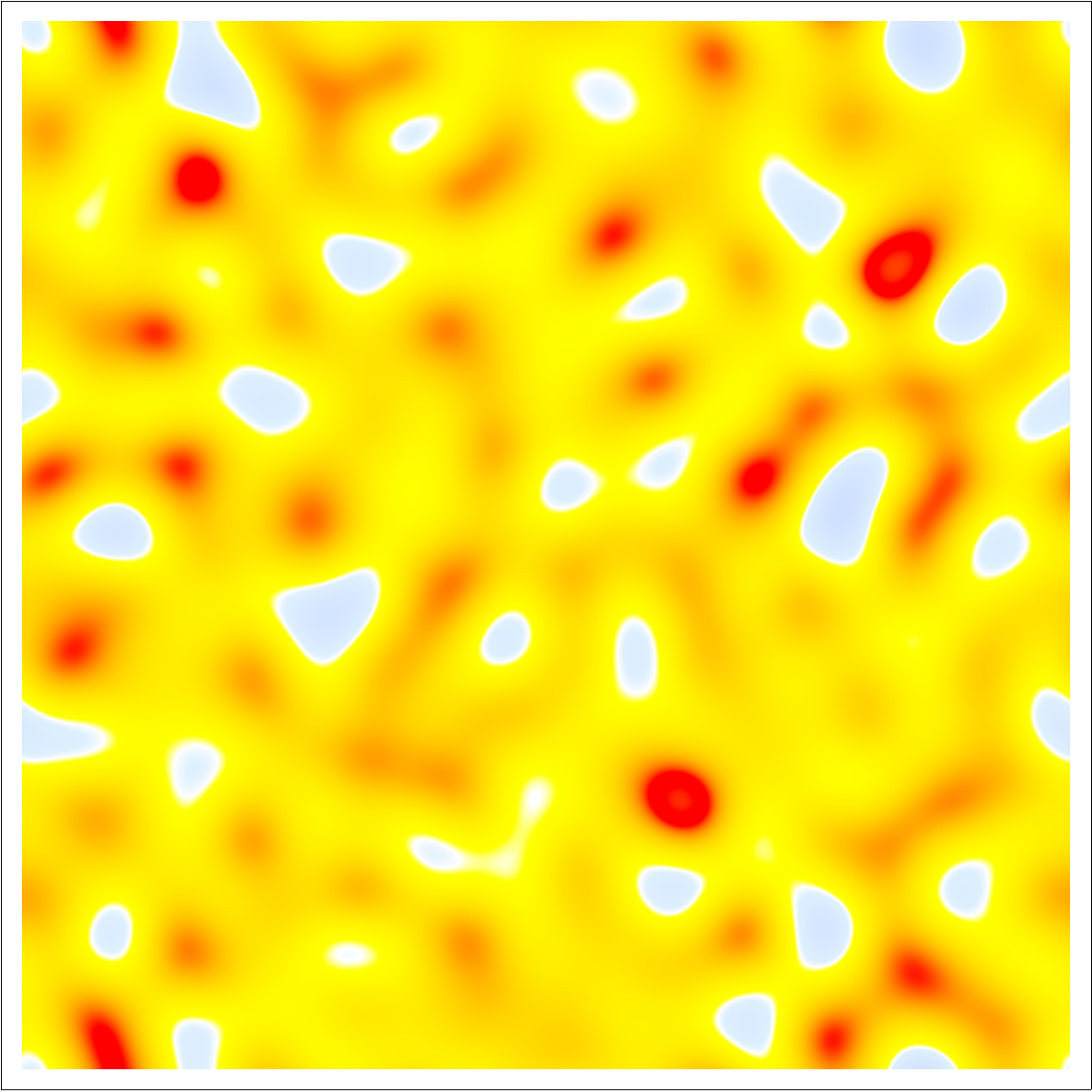} 
\includegraphics[width=0.33\textwidth]{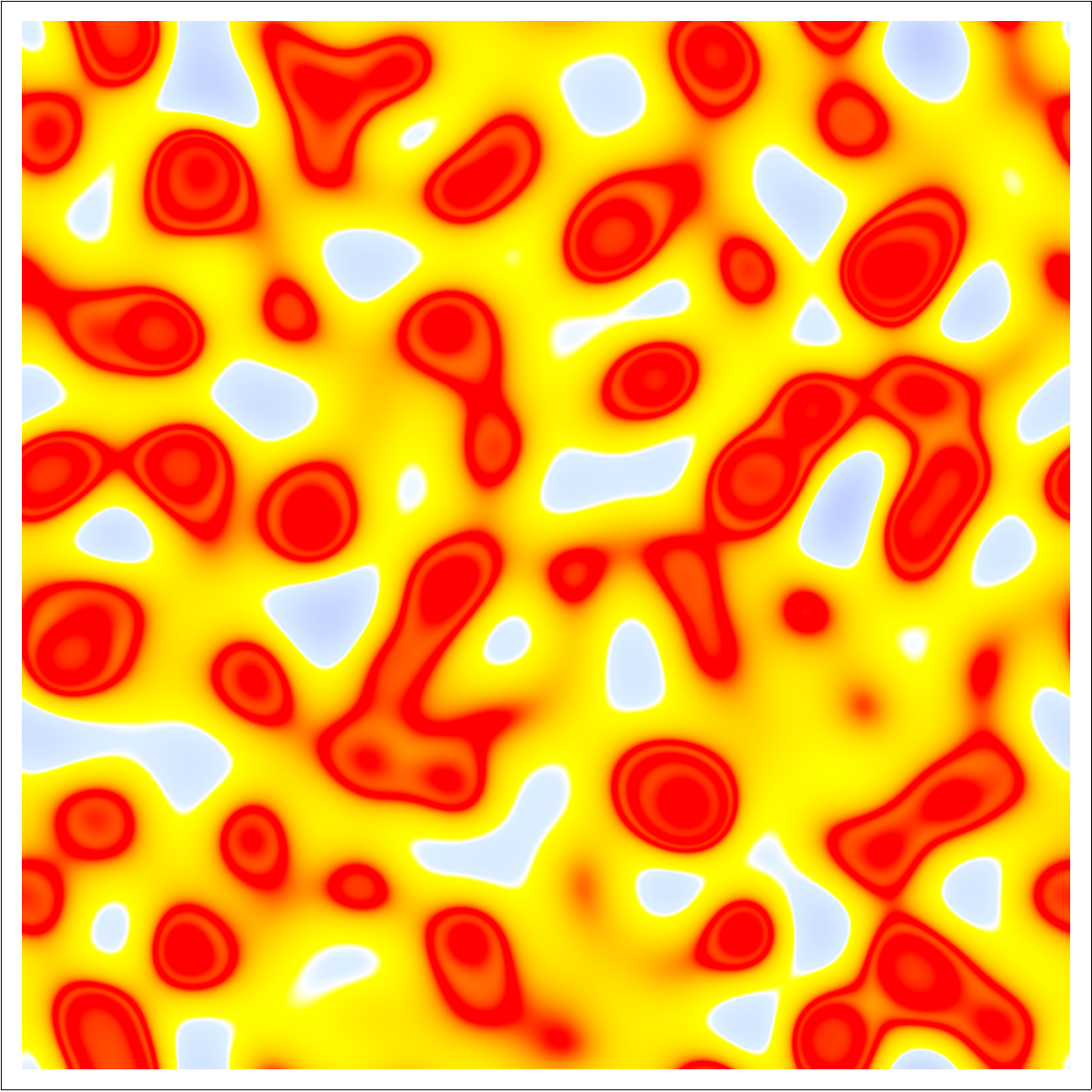}\\
\includegraphics[width=0.33\textwidth]{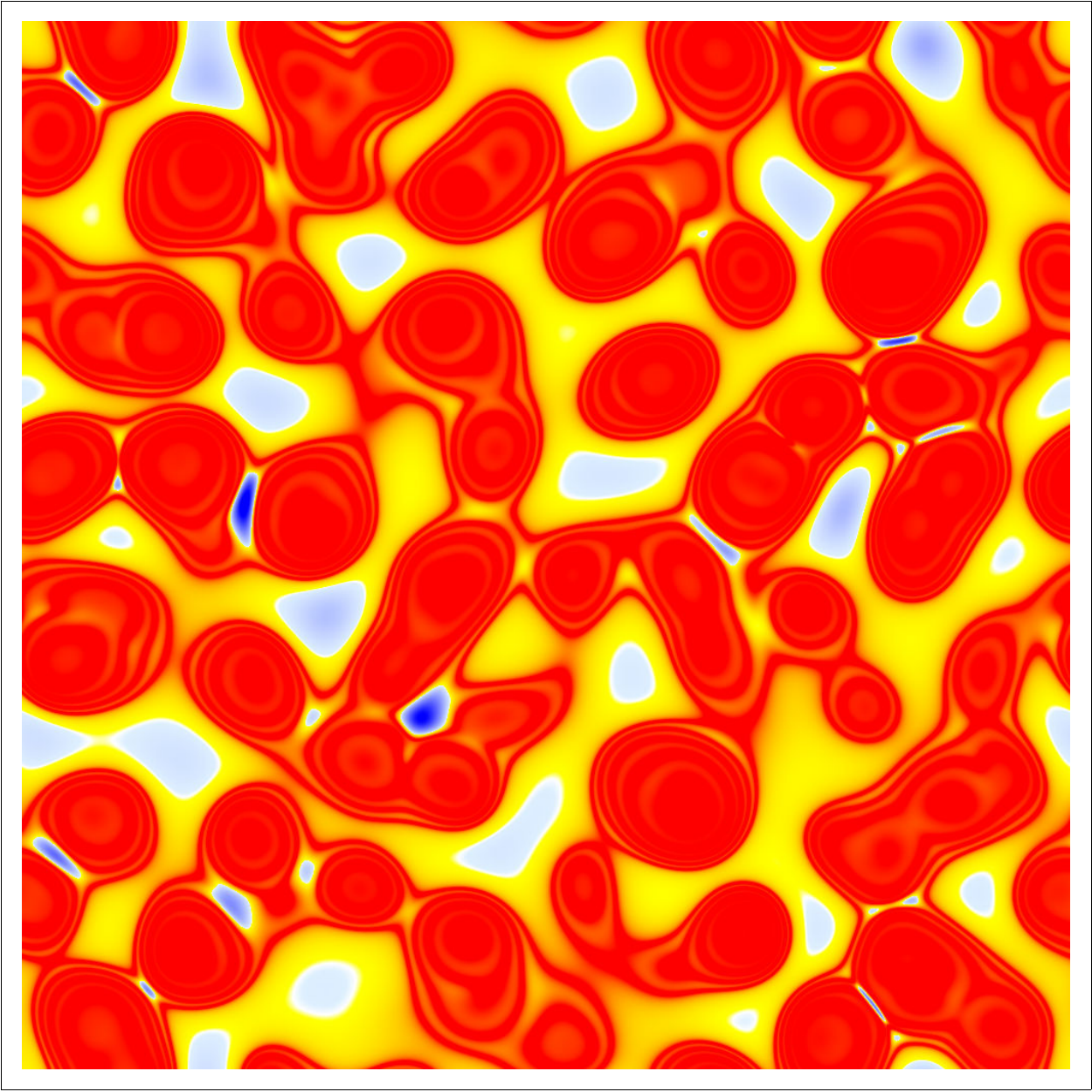} 
\includegraphics[width=0.33\textwidth]{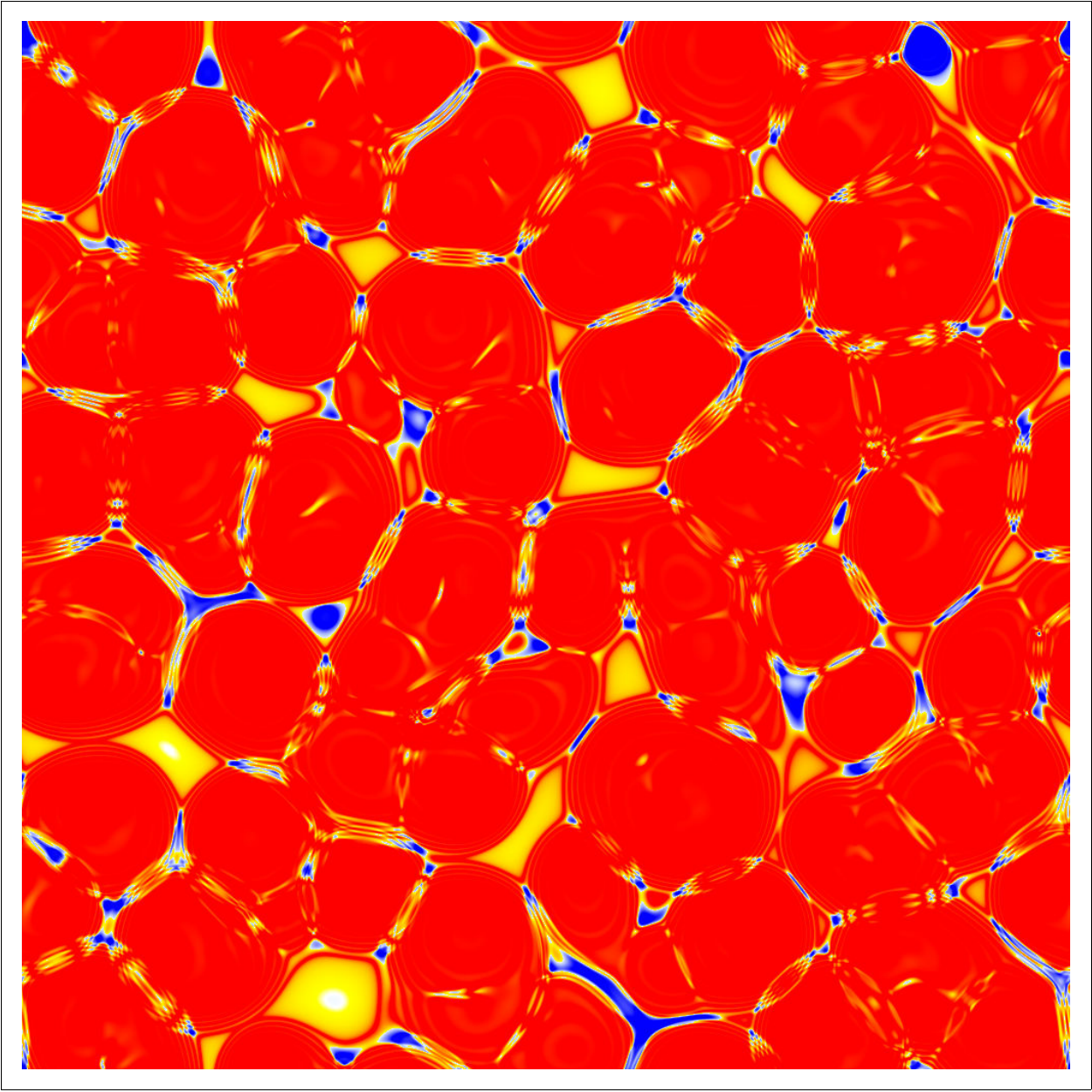} 
\includegraphics[width=0.33\textwidth]{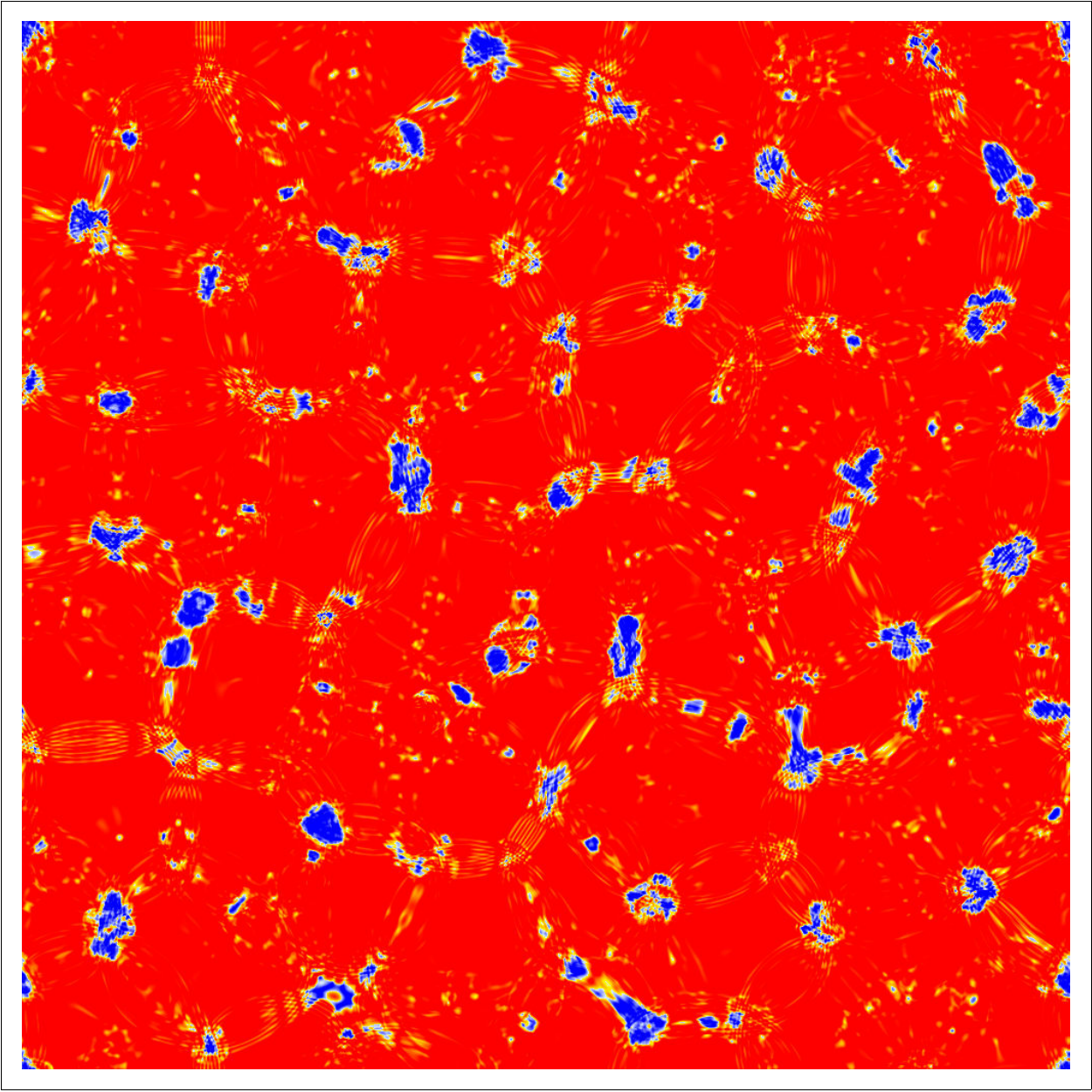}\\
\includegraphics[width=0.33\textwidth]{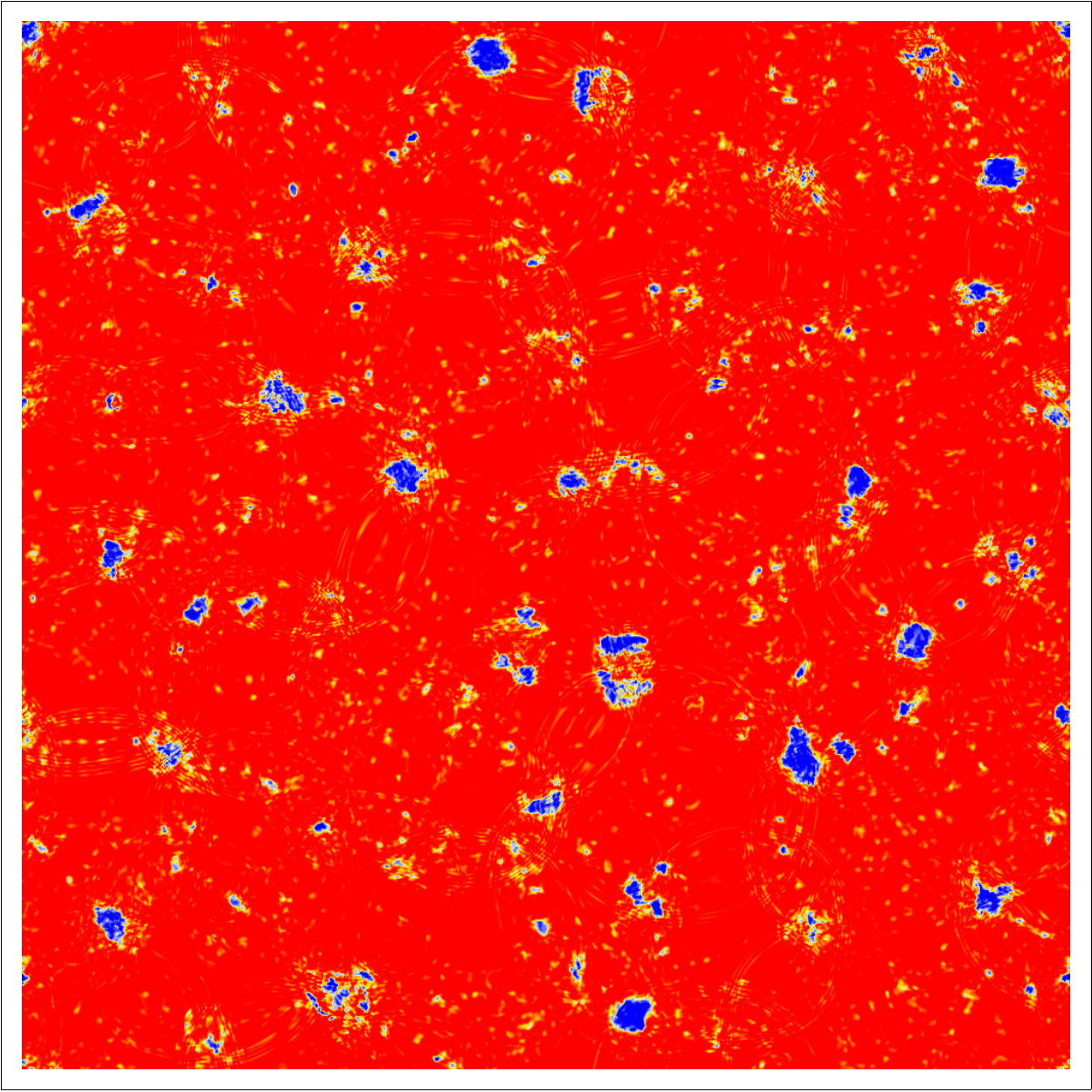}
\includegraphics[width=0.33\textwidth]{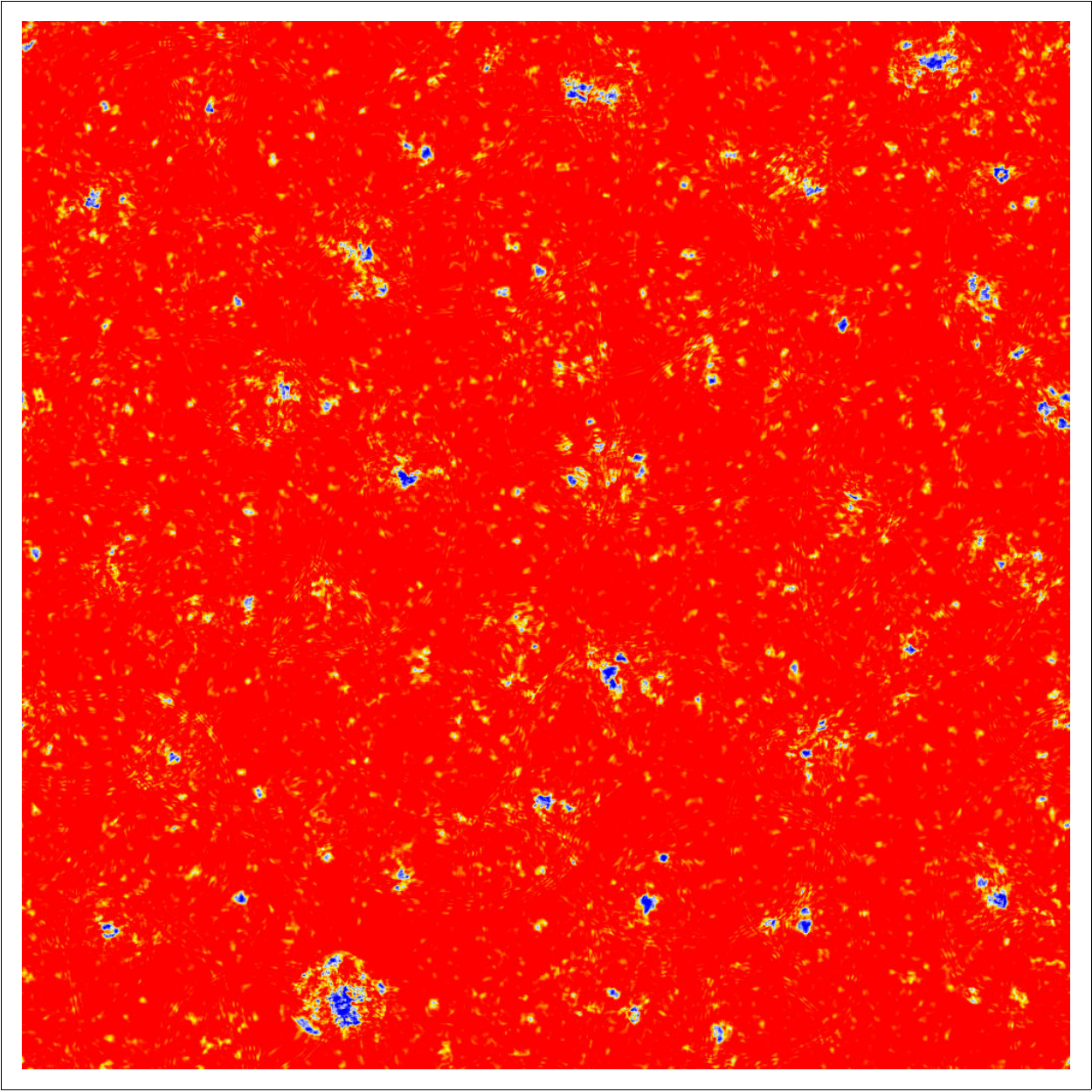} 
\includegraphics[width=0.33\textwidth]{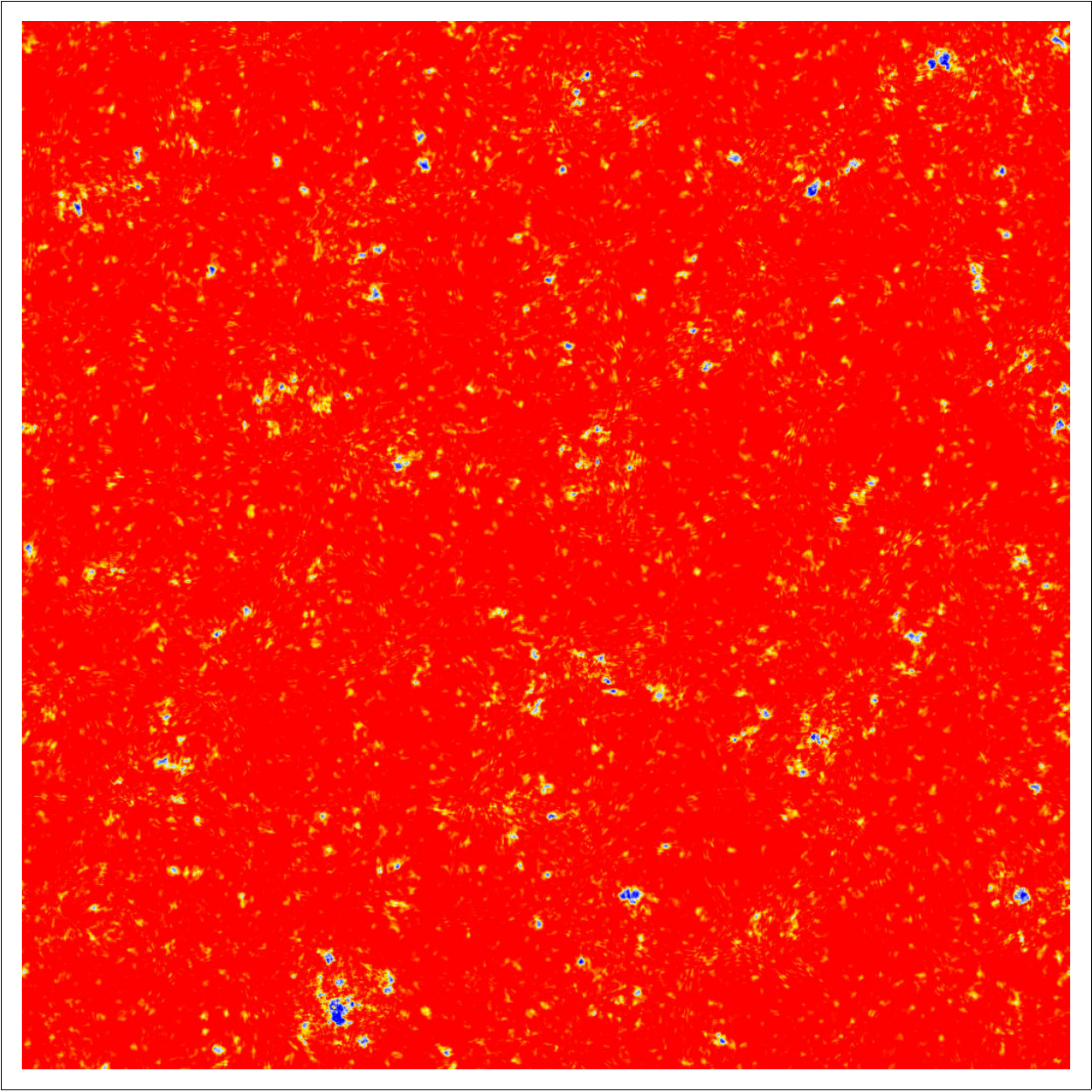}\\
\includegraphics[width=0.75\textwidth]{graphics/temp.pdf}
\end{array}$
  \caption{$\phi(x)$ from the lattice simulation at times $25.55$, $26$, $27.1$, $28.1$, $29.55$, $31.56$, $33$, $35$, $37$ in  units of $k_{\rm peak}^{-1}$. The corresponding ratios of points with $\phi<0$ can be read off fig.~\ref{fig:v2}, where the slices correspond to the blue dots in the lower right plot. The legend is shown in units of $v$. The simulation clearly shows the initial overshooting, formation and dissipation of a filament structure, and eventually localized bubbles with $\phi < 0$ which decay over time, as discussed in section~\ref{sec:latticeSlices}. Many bubbles also oscillate between $\phi \sim -v$ and $\phi \sim v$ (especially the smaller ones), though their oscillation is out of phase and thus only visible with a higher time resolution. A higher time resolution movie of the evolution can be downloaded at \cite{baselAnimation}.}
  \label{fig:v5slices}
\end{figure}

\begin{figure}[hp]
  \centering
\text{$v=10^{-2}\mpl$,$\;\;$ $N=4096$,$\;\;$ $D=2$, $\;\;$ $k_{\rm uv} = 100k_{\rm peak}$.}\\
\includegraphics[width=0.95\textwidth]{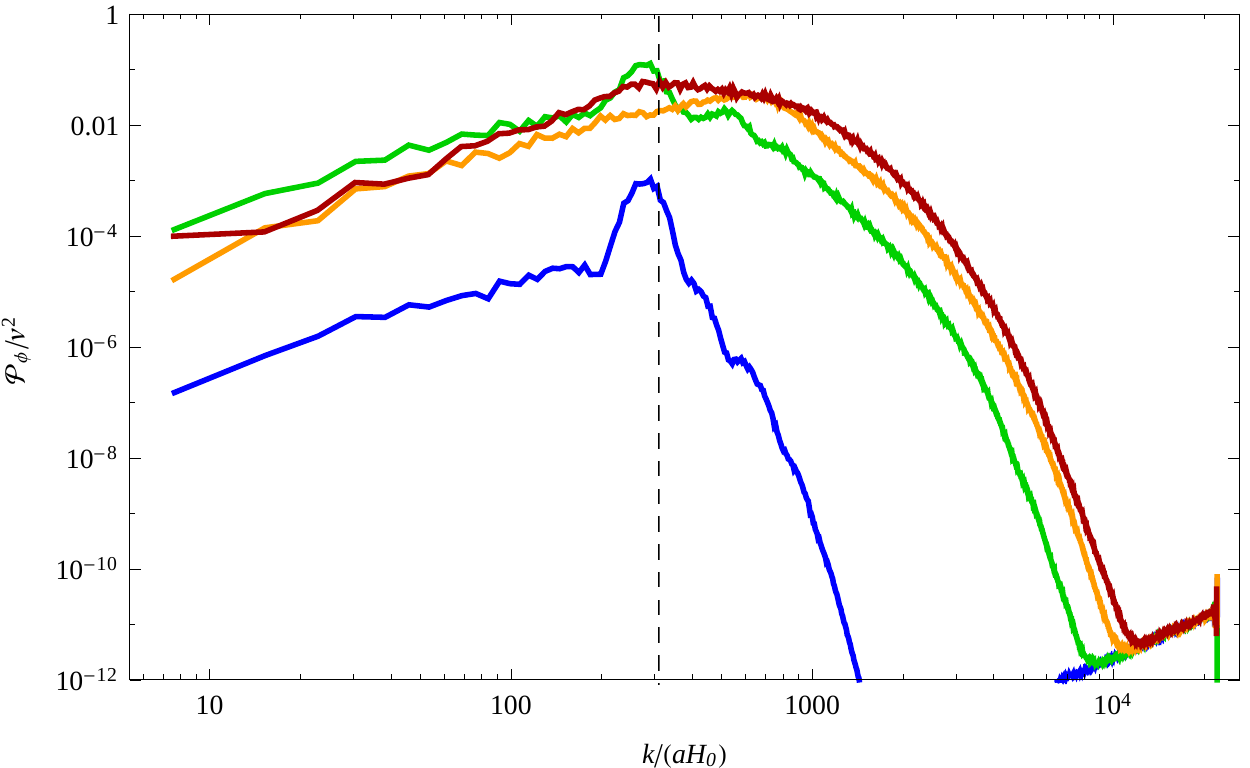}
\centering
\text{$v=10^{-5}\mpl$,$\;\;$ $N=4096$,$\;\;$ $D=2$, $\;\;$ $k_{\rm uv} = 100k_{\rm peak}$.}\\
\includegraphics[width=0.95\textwidth]{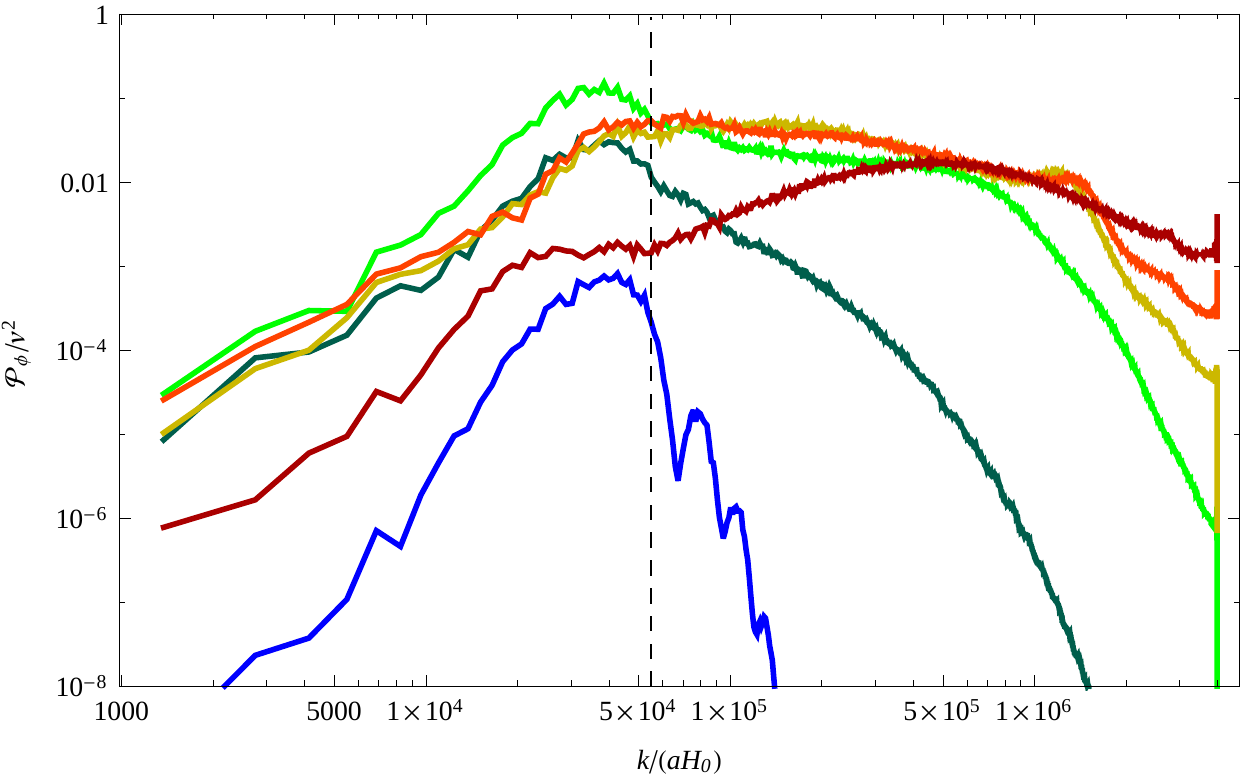}
  \caption{The spectrum $\mathcal{P}_{\phi}$, defined by $\langle\delta\phi^2\rangle\equiv\int (d\ln k)\mathcal{P}_{\phi}$. \newline
For $v=10^{-2}\mpl$ (above), the spectra are shown at times $40$, $44$, $47$ and $60$ (blue to red) in units of $k_{\rm peak}^{-1}$. The spectra are progressively shifted towards the ultraviolet, although the tail is still dominated by the vacuum fluctuations at the end of the simulation. We conclude that the relevant scales are well resolved throughout the entire simulation. \newline
For $v=10^{-5}\mpl$, the spectra are shown at times $25$, $26$, $28$, $30$, $32$ and $40$ (blue to red) in units of $k_{\rm peak}^{-1}$. The ultraviolet tail progressively grows and by the end of the simulation the peak of the spectrum is only one order of magnitude larger than the ultraviolet tail, which limits our ability to study the late-time behaviour of the overshooting bubbles. However, during most of the evolution the neglected UV power is much smaller than the power at the well-resolved scales around $k_{\rm peak}$.}
  \label{fig:v2spectra}
\end{figure}

Figs.~\ref{fig:v2slices}--\ref{fig:v5slices} show the field values on the 2D lattice at 9 different points in time, for $v=10^{-2}\mpl$ and $v=10^{-5}\mpl$, starting around the time when the initial hill crossing occurs.\footnote{Note that, to minimize image sizes, the slices shown in figs.~\ref{fig:v2slices}--\ref{fig:v5slices} are from a simulation with $N=1024$, compared to $N=4096$ of figs.~\ref{fig:v2}--\ref{fig:v2spectra}. Nevertheless, the initial fluctuations are the same in both simulations and the power lost in the infrared is subdominant to that at the peak scale $k_{\rm peak}$, as one can see from fig.~\ref{fig:v2spectra}.} The red-coloured regions have field values around the positive minimum $\phi \sim v$, the yellow and light blue regions are close to the hilltop $\phi \sim 0$, and the dark blue regions are around the negative minimum $\phi \sim -v$. Animated movies showing many more time steps for both simulations are available at \cite{baselAnimation}.

We clearly see that overshooting to $\phi < 0$ indeed occurs, creating small negative regions with $\phi < 0$ which are separated by a distance of roughly $\lambda_{\rm peak} = 2\pi/k_{\rm peak}$. The size $d_-$ of these negative regions is significantly smaller than the typical distance $\lambda_{\rm peak}$, especially for $v=10^{-2}\mpl$. This is what one would expect from overshooting due to fluctuations dominated by a specific wavelength $\lambda_{\rm peak}$: the regions which overshoot are those close to the maxima of this wave, which explains both their small size $d_- < \lambda_{\rm peak}$ and the typical distance $\lambda_{\rm peak}$.

During the early stage around the initial overshooting, the dynamics is very different for $v=10^{-2}\mpl$ and $v = 10^{-5}\mpl$. For $v=10^{-2}\mpl$, the initial overshooting occurs while $\braket{\phi} \sim v$. The initial negative regions are small, and they very quickly start oscillating towards $\phi \sim -v$. For $v=10^{-5}\mpl$, the initial overshooting occurs close to the hilltop with $\braket{\phi} \ll v$, with larger initial size of the negative regions. Although the first three steps in fig.~\ref{fig:v5slices} look similar to the early stages of the formation of a domain wall network, the subsequent evolution is qualitatively different (best seen in the animated movie \cite{baselAnimation}) and instead leads to many isolated bubbles with $\phi \sim -v$ completely enveloped in a single connected region with $\phi \sim v$.

In both cases, we eventually end up with small overshooting bubbles (blue), roughly separated by the distance $\lambda_{\rm peak}$, which are connected by filaments in which the field is close to the hilltop (yellow). These filaments then vanish, and only the blue overshooting bubbles remain.

For $v = 10^{-2}\mpl$, we can clearly see that the overshooting bubbles oscillate between $\phi < 0$ and $\phi > 0$. These oscillations are localized and mostly in phase: we can clearly see that the blue regions vanish in the $5$th, $7$th and $9$th lattice slice, and most of them reappear in the $6$th and $8$th slice at approximately their original positions. Over time, the oscillations get damped and some of the blue regions evaporate. The oscillations of the different bubbles also get out of phase over time, which is why the $9$th slice has more blue bubbles than the $5$th: in the $5$th slice, all bubbles have just oscillated towards $\phi > 0$ at the same time, whereas in the $9$th slice, some bubbles are lagging behind (or ahead) and are still (or already) at $\phi < 0$.

For $v=10^{-5}\mpl$, the blue bubbles in fig.~\ref{fig:v5slices} do not clearly exhibit any coherent oscillations. The largest bubbles indeed tend to not oscillate but quickly fragment into smaller structures. However, some of the smaller blue bubbles oscillate between $\phi < 0$ and $\phi > 0$, but these oscillations are strongly out of phase so that fig.~\ref{fig:v5slices} mostly shows the overall decay of the oscillations.\footnote{The oscillations are clearly visible in the animations containing more time slices, which are available online \cite{baselAnimation}.}

We might interpret these oscillating bubbles as ``oscillons''. It has been shown in the literature that such oscillons can form from collapsing bubbles in double-well potentials, and that they can be very long-lived, depending on their properties (like radius and initial field displacement) \cite{Copeland:1995fq}. While we observe that some of the oscillon-like bubbles with $\phi < 0$ vanish within a few oscillations, our simulation does not extend to sufficiently late times to check whether the others might remain stable over a longer timescale, or whether the bubbles which disappeared continue as long-lived oscillons which are not ``hill crossing''. We note that recently, for a double-well potential similar to ours, oscillons (not ``hill crossing'' ones) have been observed which remain long-lived on cosmological timescales \cite{Gleiser:2014ipa}.

In summary, we find that hill crossing occurs and produces regions with $\phi < 0$. However, these regions do not form domains settling at $\phi = -v$ separated from the other vacuum by domain walls, but instead they behave like localized oscillations between $\phi \sim -v$ and $\phi \sim v$. While some of these oscillon-like bubbles evaporate over time, our simulation does not clearly indicate whether or not some of them may be longer-lived.

\subsubsection{Mean, variance and fraction of negative field regions}

\begin{figure}[hp]
  \centering
\text{$v=10^{-2}\mpl$,$\;\;$ $N=4096$,$\;\;$ $D=2$, $\;\;$ $k_{\rm uv} = 100k_{\rm peak}$.}\\$
\begin{array}{cc}
\includegraphics[height=5cm]{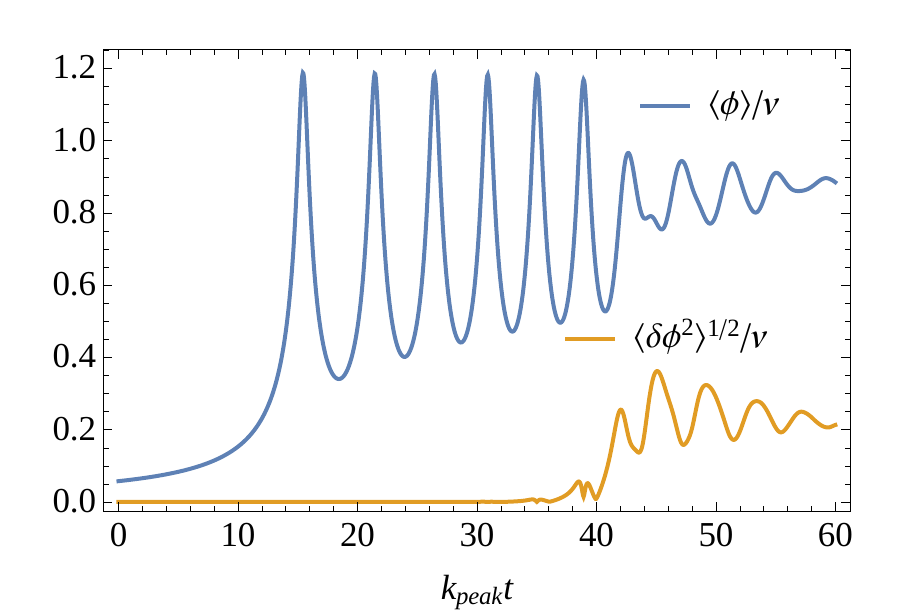}
\includegraphics[height=5cm]{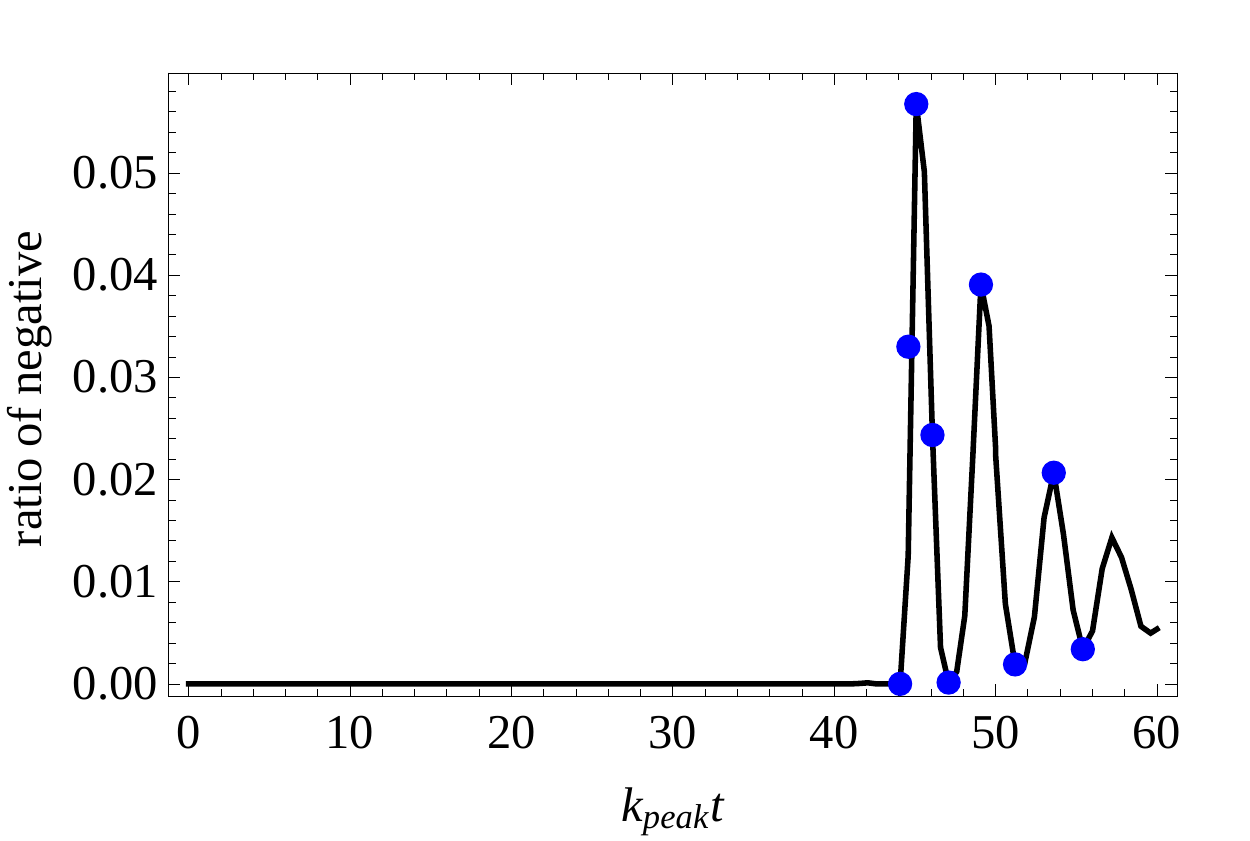}
\end{array}$
\centering
\text{$v=10^{-5}\mpl$,$\;\;$ $N=4096$,$\;\;$ $D=2$, $\;\;$ $k_{\rm uv} = 100k_{\rm peak}$.}\\$
\begin{array}{cc}
\includegraphics[height=5cm]{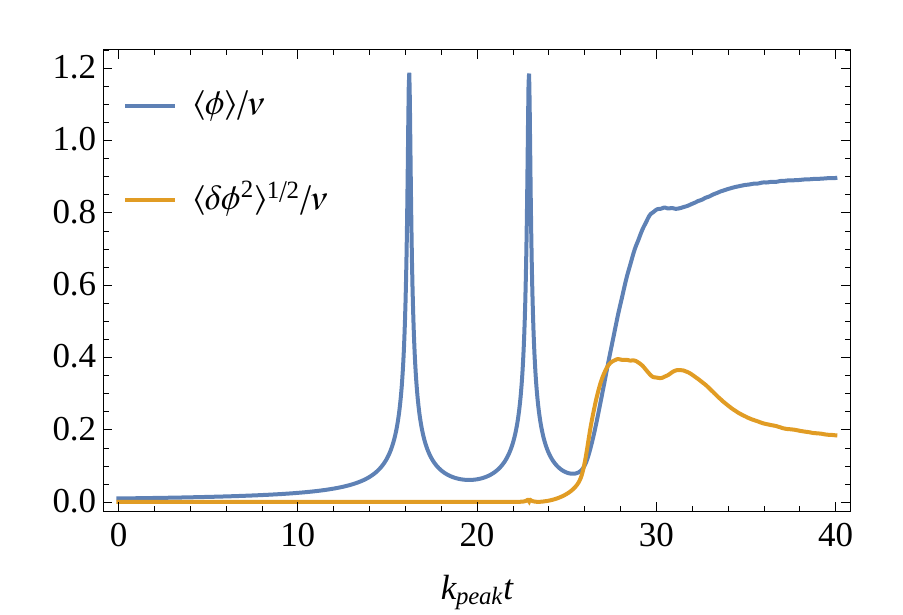}
\includegraphics[height=5cm]{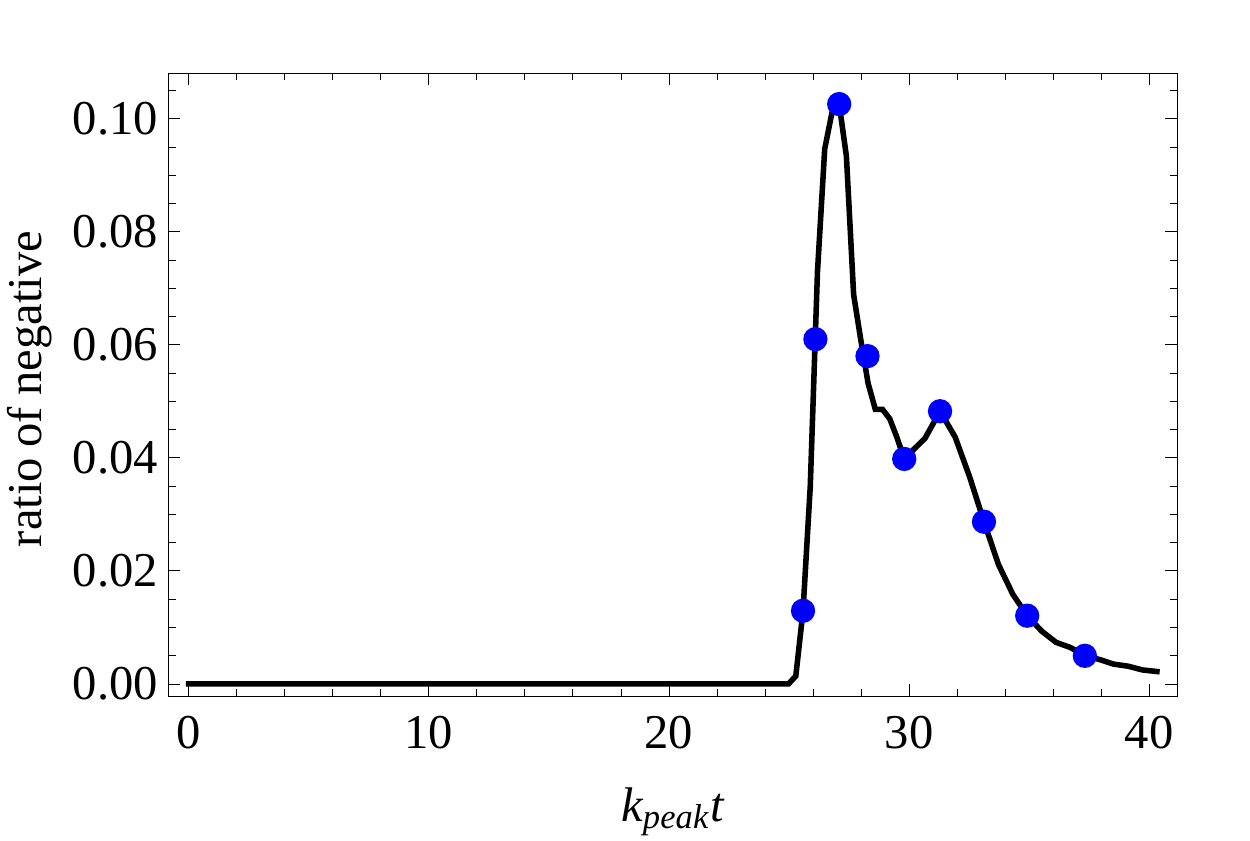} 
\end{array}$
  \caption{\textit{Left}: The variance $\sqrt{\langle\delta\phi^2\rangle}$ and the mean $\langle\phi\rangle$ in units of $v$ as a function of $k_{\rm peak} t$ for $v=10^{-2}\mpl$ (above) and $v=10^{-5}\mpl$ (below). We can clearly see how the mean performs six or two oscillations, respectively, before the variance grows to values $\braket{\delta \phi} \sim \braket{\phi}$ and non-linearities become important. This matches the expectations from the linear analysis, see figs.~\ref{fig:linearSpectra} and \ref{fig:linearSpectraWallCrossing}.
\newline
  \textit{Right}: The fraction of lattice points where $\phi(x)<0$ as a function of $k_{\rm peak} t$ for $v=10^{-2}\mpl$ (above) and $v=10^{-5}\mpl$ (below). The blue dots correspond to the slices in figs.~\ref{fig:v2slices} and \ref{fig:v5slices}.\newline We clearly see that around the onset of the non-linear stage, hill crossing happens for both values of $v$. For $v=10^{-2}\mpl$, the crossing regions oscillate (initially in phase) between $\phi \sim -v$ and $\phi \sim v$, which makes the ratio of points with $\phi < 0$ oscillate. Over time, the amplitude of this oscillation is damped, as some of these oscillating bubbles decay. They also get out of phase, which explains why the minima of the ratio of negative points get larger during the later oscillations. For $v=10^{-5}\mpl$, the oscillations are less pronounced: some of the smaller bubbles oscillate, larger bubbles tend to fragment into smaller substructures. In this case, the bubbles' oscillations are also out of phase already initially, which is why the oscillations are not clearly visible in the plots (which average over all of the many bubbles in our lattice).}
  \label{fig:v2}
\end{figure}

The left plots in fig.~\ref{fig:v2} show the mean $\braket{\phi}$ and perturbation amplitude $\sqrt{\braket{\delta \phi^2}}$ of the inflaton field. We see the first few coherent oscillations of the background inflaton field during which the perturbations remain relatively small. When the perturbation amplitude becomes comparable to the mean, the coherent background oscillations break down and hill crossing occurs. This happens during the $7$th oscillation for $v=10^{-2}\mpl$ and during the $3$rd oscillation for $v=10^{-5}\mpl$, in agreement with our expectations from the linear analysis. For $v=10^{-2}\mpl$, some coherent oscillations in $\braket{\phi}$ and $\braket{\delta \phi^2}$ remain after the hill crossing, as the negative bubbles oscillate in phase between $\phi < 0$ and $\phi > 0$. For $v=10^{-5}\mpl$, the averages $\braket{\phi}$ and $\braket{\delta \phi^2}$ do not oscillate because the individual bubbles' oscillations are out of phase and therefore roughly average out.

The hill crossing is most obvious in the right plots of fig.~\ref{fig:v2}, which show the ratio of points with $\phi < 0$ as a percentage of total points on the lattice. For $v=10^{-2}\mpl$, this plot also clearly shows both the in-phase oscillations of these points between $\phi < 0$ and $\phi > 0$, and the overall decay of the negative regions over time. One can also see that the oscillations gradually get out of phase: this is why the minima in this figure become larger during later oscillations.

For $v=10^{-5}\mpl$, the right plot of fig.~\ref{fig:v2} shows no clear oscillations because the individual bubbles oscillate out of phase. However, one can still see both the initial overshooting and the subsequent decay of the overshooting bubbles.

We also see that just after hill crossing, the ratio of points with $\phi < 0$ is larger for smaller $v$. We have observed this general tendency in several lattice runs, though the exact ratio varies with the random phases chosen when initializing the lattice, see section~\ref{sec:latticeInit}.

\subsubsection{Power spectrum}
Fig.~\ref{fig:v2spectra} depicts the power spectra for different points in time, starting from around the time of hill crossing. We have checked that until shortly before hill crossing, the power spectra in the lattice simulation match the results from the linear analysis.

Around the time of hill crossing, we can see that the non-linearities wash out the oscillatory features in the spectrum for $k > k_{\rm peak}$, and that they transfer power from $k_{\rm peak}$ to larger $k$. For example, hill crossing transfers some power by producing small overshooting regions of size $d_- < \lambda_{\rm peak}$ from perturbations of a longer wavelength $\lambda_{\rm peak}$. As explained above, this is expected because overshooting should happen near the local maxima of the dominant perturbation modes, and such regions around the local maxima are necessarily smaller than the wavelength.

For $v=10^{-2}\mpl$, the spectrum remains peaked at $k_{\rm peak}$, whereas for $v=10^{-5}\mpl$, the spectrum develops much more power at small scales $k > k_{\rm peak}$. This fits well to the lattice slices in figs.~\ref{fig:v2slices}--\ref{fig:v5slices}, where for $v=10^{-2}\mpl$ the bubbles roughly retain the size they had during the initial overshooting, whereas for $v=10^{-5}\mpl$ the bubbles tend to fragment and smaller features develop.

Eventually, the spectrum flattens over the $k$-range covered by the lattice (especially for $v = 10^{-5}\mpl$). This limits the maximum time for which our lattice calculations remain valid. Extending the lattice simulation beyond that time would require more lattice points $N$, and therefore much more computing power.

\subsection{Results of lattice simulations in $3$ dimensions}
\label{sec:3D}

In order to confirm the validity of our results in 3D, we repeated the simulations for $v=10^{-2}\mpl$ and $v=10^{-5}\mpl$ with $N=256^3$ points. Because of the lower resolution, we decreased the ultraviolet cutoff to $k_{\rm uv} = 50k_{\rm peak}$. The infrared cutoff is therefore $k_{\rm ir} \simeq 0.2k_{\rm peak}$. The initial conditions for the field mean and fluctuations are the same as in the 2D simulations, see table~\ref{tab:ini} and fig.~\ref{fig:spectrain}.

The 3D simulations show the same qualitative behaviour as the 2D simulations. Hill crossing happens shortly after the end of the linear phase, and we reproduce fig.~\ref{fig:v2} throughout the coherent oscillations and the initial overshooting phases. For $v=10^{-2}\mpl$, the 2D results are reproduced until the end of the simulation at $t=60k_{\rm peak}^{-1}$. On the other hand, for $v=10^{-5}\mpl$, the simulation is only reliable until $t\sim35k_{\rm peak}^{-1}$, because afterwards the spectrum is dominated by the smallest scale fluctuations in the lattice.

\begin{figure}[hp]
  \centering
\text{$v=10^{-2}\mpl$,$\;\;$ $N=256$,$\;\;$ $D=3$, $\;\;$ $k_{\rm uv} = 50k_{\rm peak}$.}\\$
\begin{array}{cc}
\includegraphics[width=0.48\textwidth]{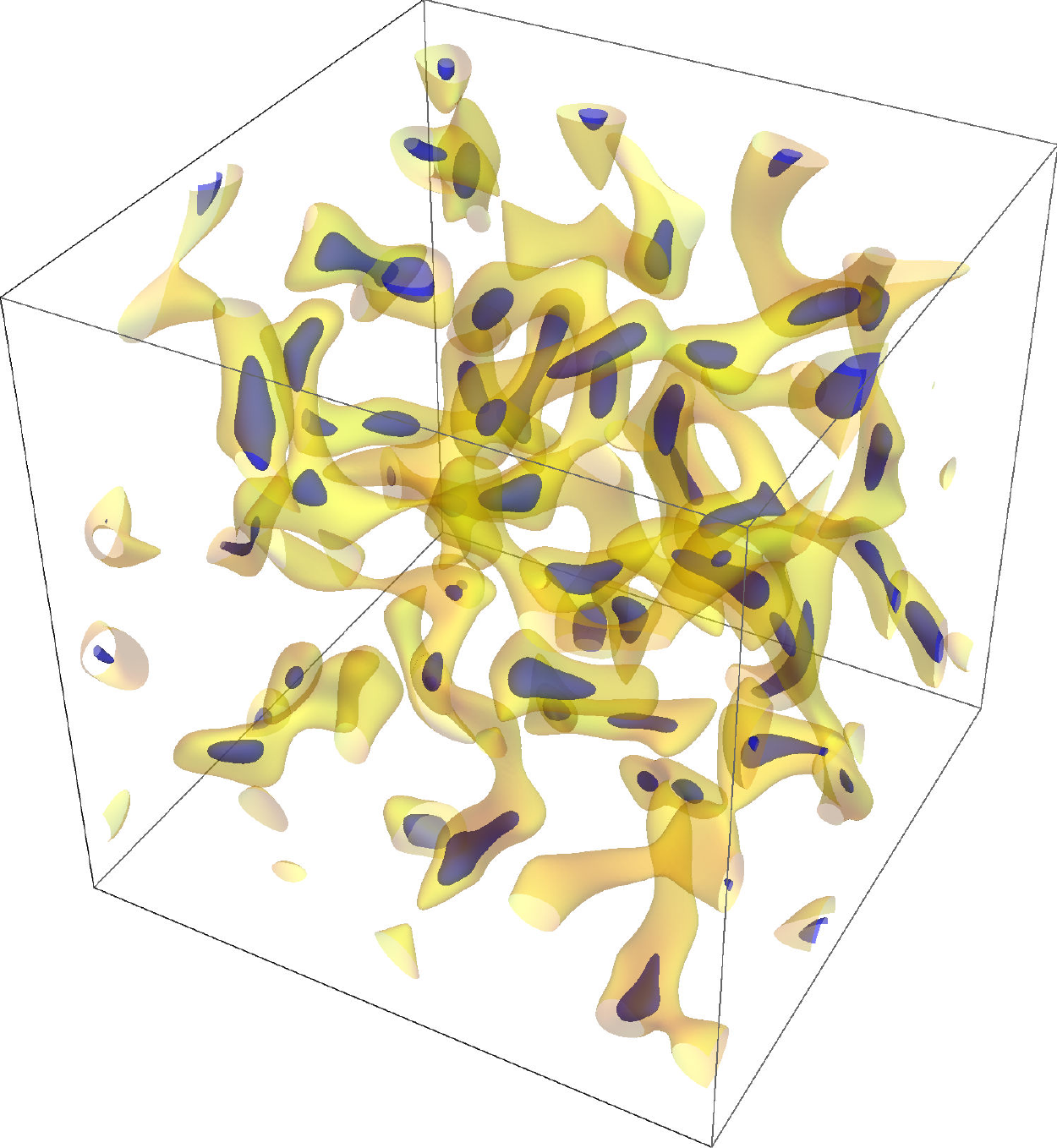}
\includegraphics[width=0.48\textwidth]{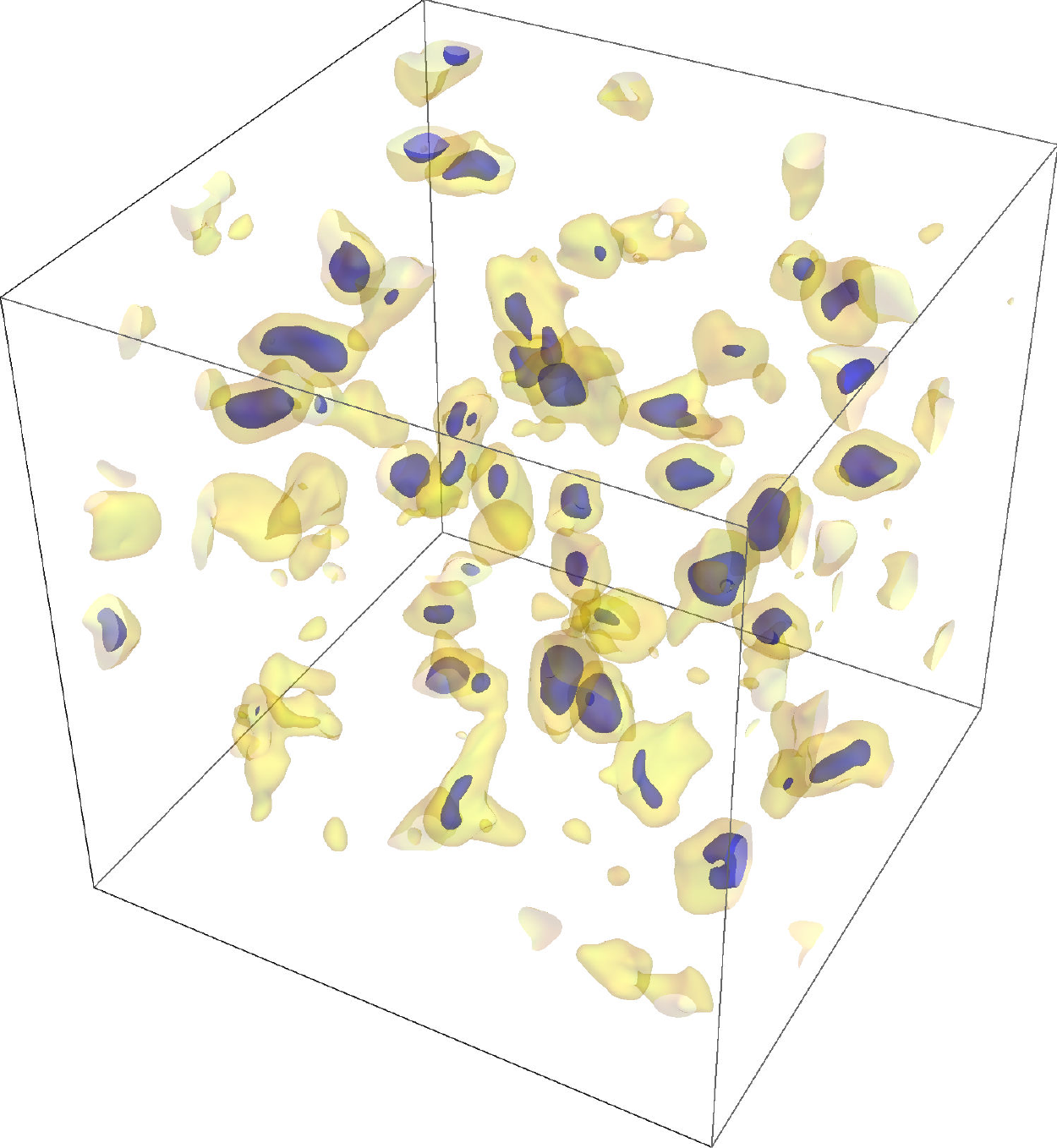}
\end{array}$
  \centering
\text{$v=10^{-5}\mpl$,$\;\;$ $N=256$,$\;\;$ $D=3$, $\;\;$ $k_{\rm uv} = 50k_{\rm peak}$.}\\$
\begin{array}{cc}
\includegraphics[width=0.48\textwidth]{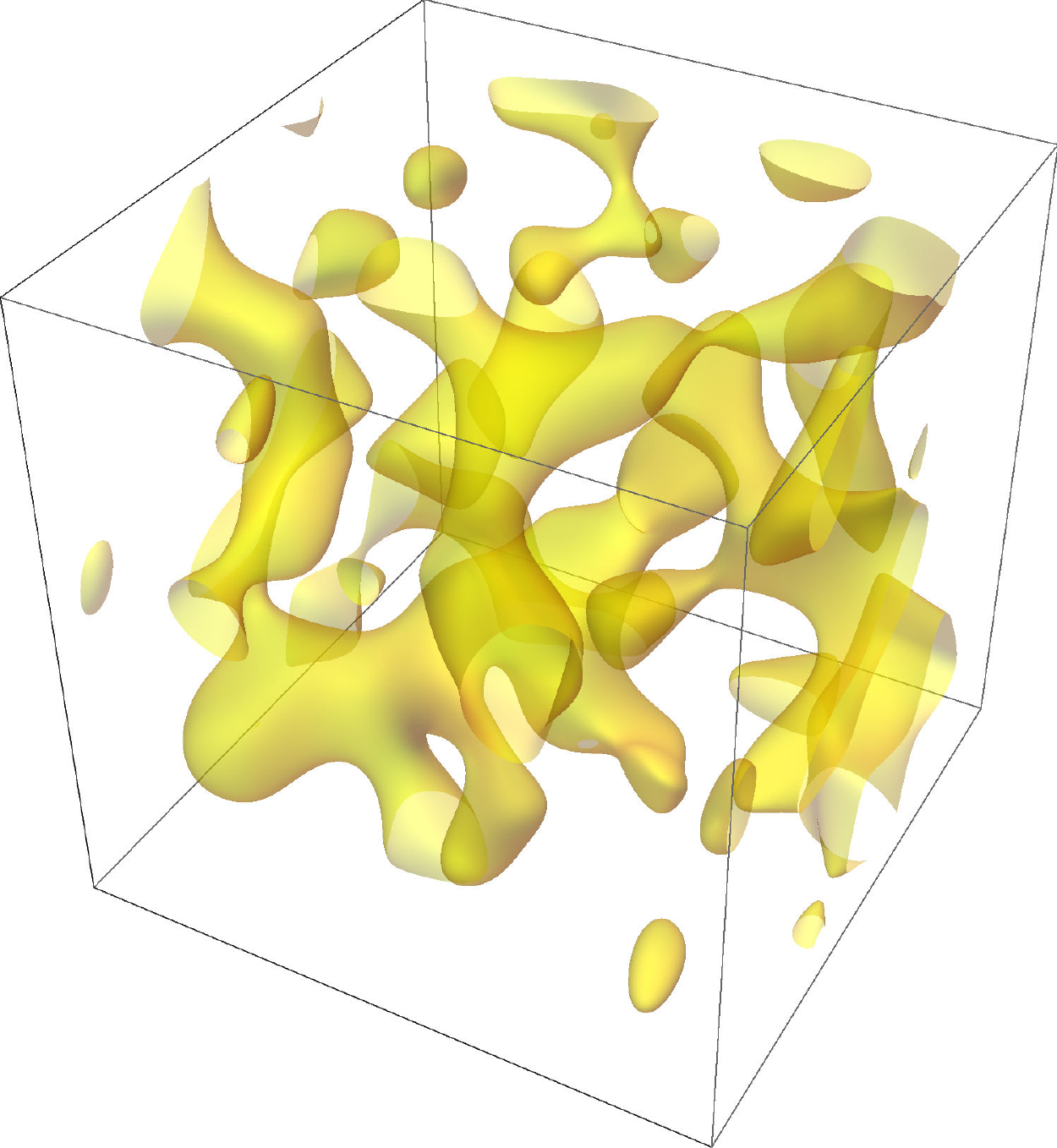}
\includegraphics[width=0.48\textwidth]{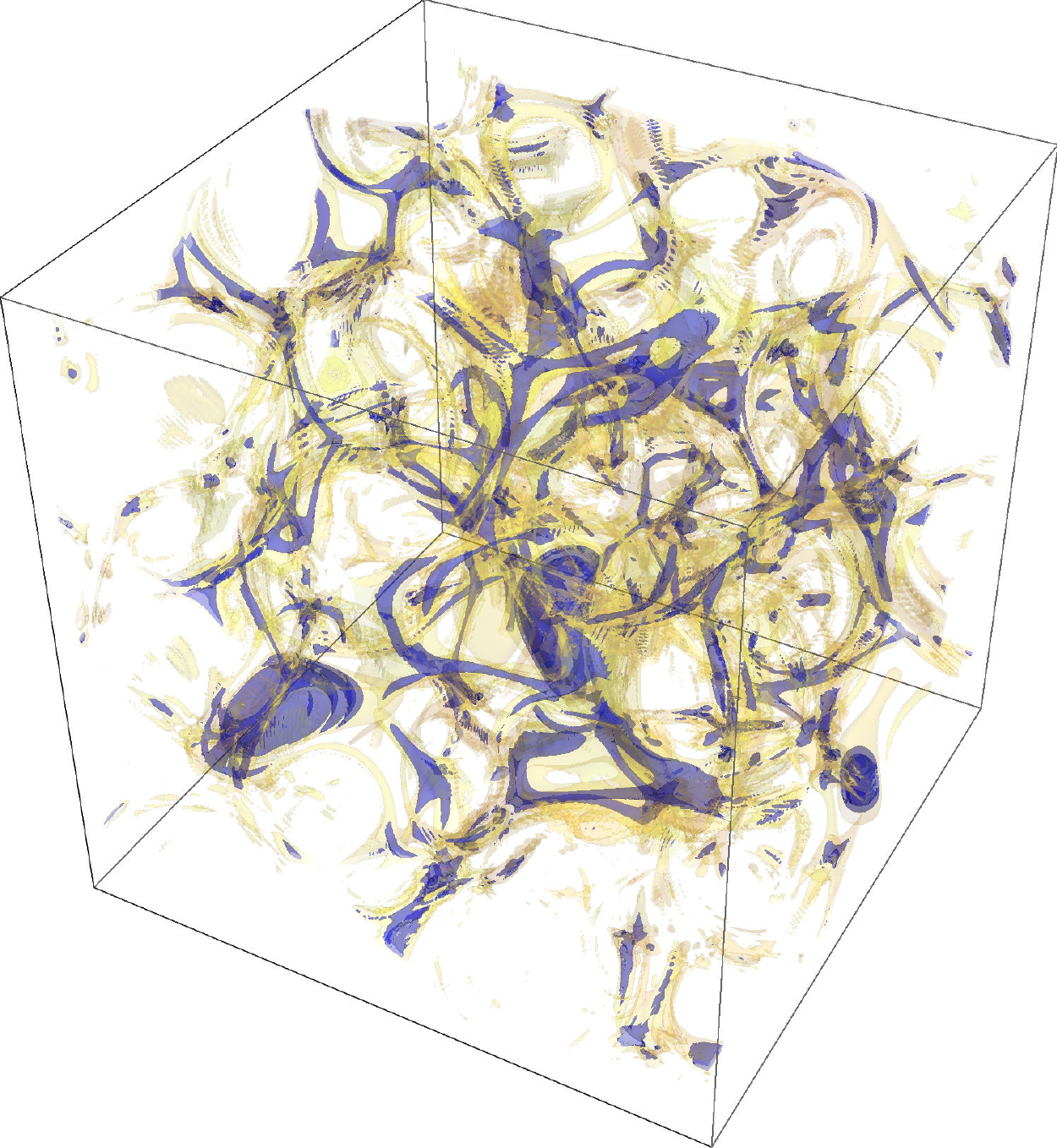}
\end{array}$
  \caption{$\phi(x)$ at fixed times from the 3D lattice simulations for $v=10^{-2}\mpl$ (above) and $v=10^{-5}\mpl$ (below). The contours correspond to points where $\phi=0$ (yellow) and $\phi=-0.8v$ (blue).\newline
 For $v=10^{-2}\mpl$, the left box corresponds to $t=45.3 k_{\rm peak}^{-1}$ and $5.8\%$ of points with $\phi<0$, and the right box to $t=53.7 k_{\rm peak}^{-1}$ and $3.2\%$ of points with $\phi<0$. The initial filament-shaped negative regions evolve into bubbles with a tendency towards a spherical shape. As in the 2D simulations, these bubbles perform localized oscillations between $\phi \sim -v$ and $\phi \sim v$. The oscillations are visible in the animation containing more time slices, available online \cite{baselAnimation}. \newline
 For $v=10^{-5}\mpl$, the left box corresponds to $t=26.7 k_{\rm peak}^{-1}$ and $10.7\%$ of points with $\phi<0$, and the right box to $t=30 k_{\rm peak}^{-1}$ and $4.4\%$ of points with $\phi<0$. The initial filament-shaped negative regions are larger than for $v=10^{-2}\mpl$. They fragment to smaller structures in a manner similar to our 2D simulations; the right box matches the corresponding 2D result, which is roughly the middle slice in fig.~\ref{fig:v5slices}. However, due to the limited resolution in 3D we cannot resolve the subsequent evolution well enough to ascertain that they form oscillating bubbles as they did in 2D.}
  \label{fig:3D}
\end{figure}

Time slices of the 3D simulations are shown in fig.~\ref{fig:3D}.\footnote{An animation of the 3D simulation for $v=10^{-2}\mpl$ showing more time slices is available online \cite{baselAnimation}.} For both $v/\mpl=10^{-2}$ and $10^{-5}$, the initial hill crossing phase is characterized by filament-shaped regions where $\phi<0$. Eventually the filaments disappear and are replaced by localized oscillations between $\phi \sim -v$ and $\phi \sim v$, with typical separation $\lambda_{\rm peak} = 2\pi/k_{\rm peak}$, as we saw in the 2D simulations.

Furthermore, for $v=10^{-2}\mpl$, we can see a tendency of most overshooting regions to develop towards a spherical shape. For $v=10^{-5}\mpl$, the overshooting regions are too small for their shape to be accurately resolved in the lattice in 3D. 

Recapitulating, the 3D simulations confirm the results of the 2D simulations. The linear and initial hill crossing phases are accurately reproduced for both $v/\mpl=10^{-2}$ and $10^{-5}$. For $v=10^{-2}\mpl$ the 3D simulation remains well-resolved until the final time of the 2D simulation and we see the overshooting regions form and tend towards a spherical shape. For $v=10^{-5}\mpl$, the resolution becomes insufficient shortly after the initial hill crossing.

\section{Summary and conclusions}
\label{sec:summary}

In this paper, we have studied preheating after ``hilltop inflation'', where inflation takes place when the inflaton field rolls slowly from close to a maximum of its potential (i.e.\ the ``hilltop'') towards its minimum. When the inflaton potential is associated with a phase transition, possible topological defects produced during this phase transition, such as domain walls, are efficiently diluted during inflation. It is typically assumed that they also do not reform after inflation, i.e.\ that the inflaton field stays on its side of the ``hill'' (where without loss of generality $\phi > 0$), finally performing damped oscillations around the minimum of the potential. 

To investigate the preheating stage, we first integrated the mode equations during the linear phase of preheating, during which inhomogeneities are still small enough to be described as small perturbations on a homogeneous background. When the inhomogeneities get larger, i.e.\ during the subsequent non-linear stage, we solved the inflaton field's classical equations of motion with lattice simulations in 2D and 3D. We were particularly interested in the possibility that large inflaton fluctuations during preheating might push the inflaton field over the local maximum of the scalar potential towards the ``wrong'' vacuum, which could in principle result in the formation of domain walls. 

We found that for  $10^{-5}\lesssim v/\mpl \lesssim 10^{-2}$ the fluctuations of the inflaton field during the tachyonic oscillation phase indeed grow strong enough to allow the inflaton field to form regions in position space where it crosses ``over the top of the hill'' towards the ``wrong vacuum'' where $\phi < 0$. The negative regions with $\phi < 0$ are separated by a distance of roughly $\lambda_{\rm peak} = 2\pi/k_{\rm peak}$, as one would expect from ``hill crossing'' due to fluctuations dominated by a specific wavelength $\lambda_{\rm peak}$. Rather than forming durable domain walls, these regions develop into localized bubbles that oscillate between $\phi \sim -v$ and $\phi \sim v$, which might be interpreted as oscillons \cite{Copeland:1995fq,Adib:2002ff,Gleiser:2003uu,Graham:2006xs,Farhi:2007wj,Amin:2011hj,Zhou:2013tsa,Gleiser:2014ipa,Lozanov:2014zfa}. These oscillon-like bubbles should be included in a careful study of preheating in hilltop inflation.

As next steps beyond the work presented here, it will be interesting to study the evolution of these oscillon-like structures in more detail, to study their fate (and lifetime) after the hill crossing phase. Furthermore, it will be interesting to study the cosmological consequences of the hill crossing phase, especially when a second ``matter field'' is included which couples to the inflaton and receives its mass dynamically from the vacuum expectation value of the inflaton field (as e.g.\ in \cite{Antusch:2014qqa}). This may lead to explosive matter particle production at the hill crossing regions, but may also trigger a faster decay of the oszillating bubbles. 
Finally, in order to derive precise predictions of explicit hilltop inflation models it will be crucial to reliably calculate the expansion history of the universe, which depends on the evolution and lifetime of the hill crossing regions.

\subsection*{Acknowledgements}
This work was supported by the Swiss National Science Foundation. The authors also thank Francesco Cefal\`{a} for helpful discussions.

\end{document}